\documentclass[12pt]{article}

\newcommand{\blind}{0}

\usepackage{amsmath,amssymb,amsthm}
\usepackage{dsfont}
\usepackage{natbib}
\usepackage{multirow, array}
\usepackage{xr}
\usepackage{enumerate}
\usepackage{url} 

\usepackage{lipsum}
\usepackage{amsfonts}
\usepackage{graphicx,psfrag,epsf}
\usepackage{epstopdf}
\usepackage{algpseudocode}
\usepackage{algorithm}

\usepackage{subcaption}

\usepackage{color}
\usepackage{verbatim}

\usepackage{booktabs}

\RequirePackage[hidelinks]{hyperref}
\usepackage[noabbrev]{cleveref}

\ifpdf
  \DeclareGraphicsExtensions{.eps,.pdf,.png,.jpg}
\else
  \DeclareGraphicsExtensions{.eps}
\fi

\newcommand{\TheTitle}{Adaptive Design of Experiments for Conservative Estimation of Excursion Sets}

\title{{\TheTitle}}

\if0\blind
{
	\title{\bf \TheTitle}
	\author{
		Dario Azzimonti\footnotemark[1] \footnotemark[2], David Ginsbourger\footnotemark[3] \footnotemark[4], Cl\'ement Chevalier\footnotemark[5], \\ Julien Bect\footnotemark[6], Yann Richet\footnotemark[7]
	}
} \fi

\if1\blind
{
	\title{\bf \TheTitle}
} \fi

\usepackage{amsopn}

\newcommand{\doe}{\mathbf{X}}
\newcommand{\R}{\mathbb{R}}
\newcommand{\E}{\mathbb{E}}
\newcommand{\inSpace}{\mathbb{X}}
\newcommand{\outSpace}{\mathbb{R}}
\newcommand{\nNot}{{n}}
\newcommand{\mean}{\mathfrak{m}}
\newcommand{\covkern}{\mathfrak{K}}
\newcommand{\sd}{\mathfrak{s}}

\newcommand{\measure}{\mu}
\newcommand{\measureOf}[1][\cdot]{\measure(#1)}
\newcommand\dmeasure[1][x]{d\measure(#1)}
\newcommand{\volume}{\operatorname{vol}}
\newcommand{\volumeOf}[1][\cdot]{\volume(#1)}
\newcommand{\CE}{\operatorname{CE}}
\newcommand{\setOfInt}{\Gamma(f)}
\newcommand{\setOfIntKeff}{\Gamma(\keff)}
\newcommand{\ConsLevel}{\rho^\alpha}

\newcommand{\keff}{\operatorname{k-eff}}
\newcommand{\PuOdens}{\textrm{PuO}_2}
\newcommand{\WaterThick}{\textrm{H}_2 \textrm{O}}
\newcommand{\newEval}[1][q]{\mathbf{x}^{(#1)}}
\newcommand{\RprogLang}{\textbf{\textsf{R}}}
\newcommand{\IMSE}{\operatorname{IMSE}}
\newcommand{\tIMSE}{\operatorname{tIMSE}}

\newtheorem{proposition}{Proposition}

\newtheorem{definition}{Definition}

\theoremstyle{remark}

\begin{document}

\def\spacingset#1{\renewcommand{\baselinestretch}%
	{#1}\small\normalsize} \spacingset{1}

\if0\blind
{	
	\renewcommand{\thefootnote}{\fnsymbol{footnote}}
	\footnotetext[1]{The first author gratefully acknowledges the Swiss National Science Foundation, grant number 146354 and 167199 and the Hasler Foundation, grant number 16065. The authors thank Michael McCourt for his question on model-free comparisons at the NIPS 2017 workshop on Bayesian Optimization.}
	\footnotetext[2]{\footnotesize Istituto Dalle Molle di studi sull'Intelligenza Artificiale (IDSIA), Scuola universitaria professionale della Svizzera italiana (SUPSI), Universit\`a della Svizzera italiana (USI), Via Cantonale 2c, 6928 Manno, Switzerland}
	\footnotetext[3]{Uncertainty Quantification and Optimal Design group, Idiap Research Institute, Centre du Parc, Rue Marconi 19, PO Box 592, 1920 Martigny, Switzerland.}
	\footnotetext[4]{IMSV, Department of Mathematics and Statistics, University of Bern, Alpeneggstrasse 22, 3012 Bern, Switzerland.} 
	\footnotetext[5]{Institute of Statistics, University of Neuch\^atel, Avenue de Bellevaux 51, 2000 Neuch\^atel, Switzerland.}
	\footnotetext[6]{Laboratoire des Signaux et Syst\`emes (UMR CNRS 8506), CentraleSupelec, CNRS, Univ. Paris-Sud, Universit\'e Paris-Saclay, 91192, Gif-sur-Yvette, France.} 
	\footnotetext[7]{Institut de Radioprotection et de S\^uret\'e Nucl\'eaire (IRSN), Paris, France.} 
	
	\renewcommand{\thefootnote}{\arabic{footnote}}
		
}\fi	

	\maketitle
	
	\begin{abstract}
		We consider the problem of estimating the set of all inputs that leads a system to some particular behavior. The system is modeled by an expensive-to-evaluate function, such as a computer experiment, and we are interested in its excursion set, i.e.\ the set of points where the function takes values above or below some prescribed threshold. The objective function is emulated with a Gaussian Process (GP) model based on an initial design of experiments enriched with evaluation results at (batch-) sequentially determined input points.  
		The GP model provides conservative estimates for the excursion set, which control false positives while minimizing false negatives. We introduce adaptive strategies that sequentially select new evaluations of the function by reducing the uncertainty on conservative estimates. Following the Stepwise Uncertainty Reduction approach we obtain new evaluations by minimizing adapted criteria. Tractable formulae for the conservative criteria are derived, which allow more convenient optimization. The method is benchmarked on random functions generated under the model assumptions in different scenarios of noise and batch size. We then apply it to a reliability engineering test case. Overall, the proposed strategy of minimizing false negatives in conservative estimation achieves competitive performance both in terms of model-based and model-free indicators.
	\end{abstract}

\noindent%
\vfill

\newpage
\spacingset{1.45} 

\section{Introduction}
\label{sec:Introduction}

The problem of estimating the set of inputs that leads a system to a particular behavior is common in many applications, such as reliability engineering~\citep[see, e.g.,][]{Bect.etal2012,Chevalier.etal2014}, climatology~\citep[see, e.g.,][]{frenchSain2013spatio,BolinLindgren2014} and many other fields~\citep[see, e.g.,][]{bayarri.etal2009using,Arnaud.etal2010}. 
Here we consider a system modeled as a continuous, expensive-to-evaluate function~$f: \inSpace \rightarrow \outSpace$, where $\inSpace$ is a compact subset of $\R^d$. \Cref{sec:IRSNtestCase} shows an example of such systems. Given a few evaluations of $f$ and a fixed closed set $T \subset \outSpace$, we are interested in estimating
\begin{equation}
\setOfInt =\{ x \in \inSpace : f(x) \in T\}.
\label{eq:setOfInt}
\end{equation}
In the motivating test case in~\cref{sec:IRSNtestCase}, for example, $\setOfInt$ represents the \emph{safe region}, i.e.\ all values of the physical parameters that lead the system of interest to a subcritical response, taking $T=(-\infty,t]$ with $t\in \R$.

There is much heterogeneity in the literature on how to name $\setOfInt$. Here we follow~\citet{Adler.Taylor2007} and we call $\setOfInt$ an \emph{excursion set}. If $T=[t, +\infty)$, with $t \in \R$, $\setOfInt$ is often also called excursion set above~$t$~\citep[see, e.g.,][]{Azais.Wschebor2009,BolinLindgren2014}, but also level set~\citep{Berkenkamp2017SafeRL}. If $T=(-\infty,t]$ the set is sometimes referenced as sojourn set~\citep{Spodarev2014} or sublevel set~\citep{Gotovos.etal2013}. 
Our work is primarily motivated by the case $T=[t,+\infty)$, however it may also applied to $T=(-\infty,t]$. 

Throughout the article, we model $f$ as a realization of a Gaussian process (GP) and, following~\citet{Sacks.etal1989} and \citet{Santner.etal2018}, we emulate $f$ with the posterior GP distribution given the available function evaluations. The posterior GP distribution can be used as building block for different estimates of $\setOfInt$, see, e.g.,~\citet{Azzimonti2016}. 

\begin{figure}
			\begin{minipage}{0.48\textwidth}
				\centering
				\includegraphics[width=\linewidth]{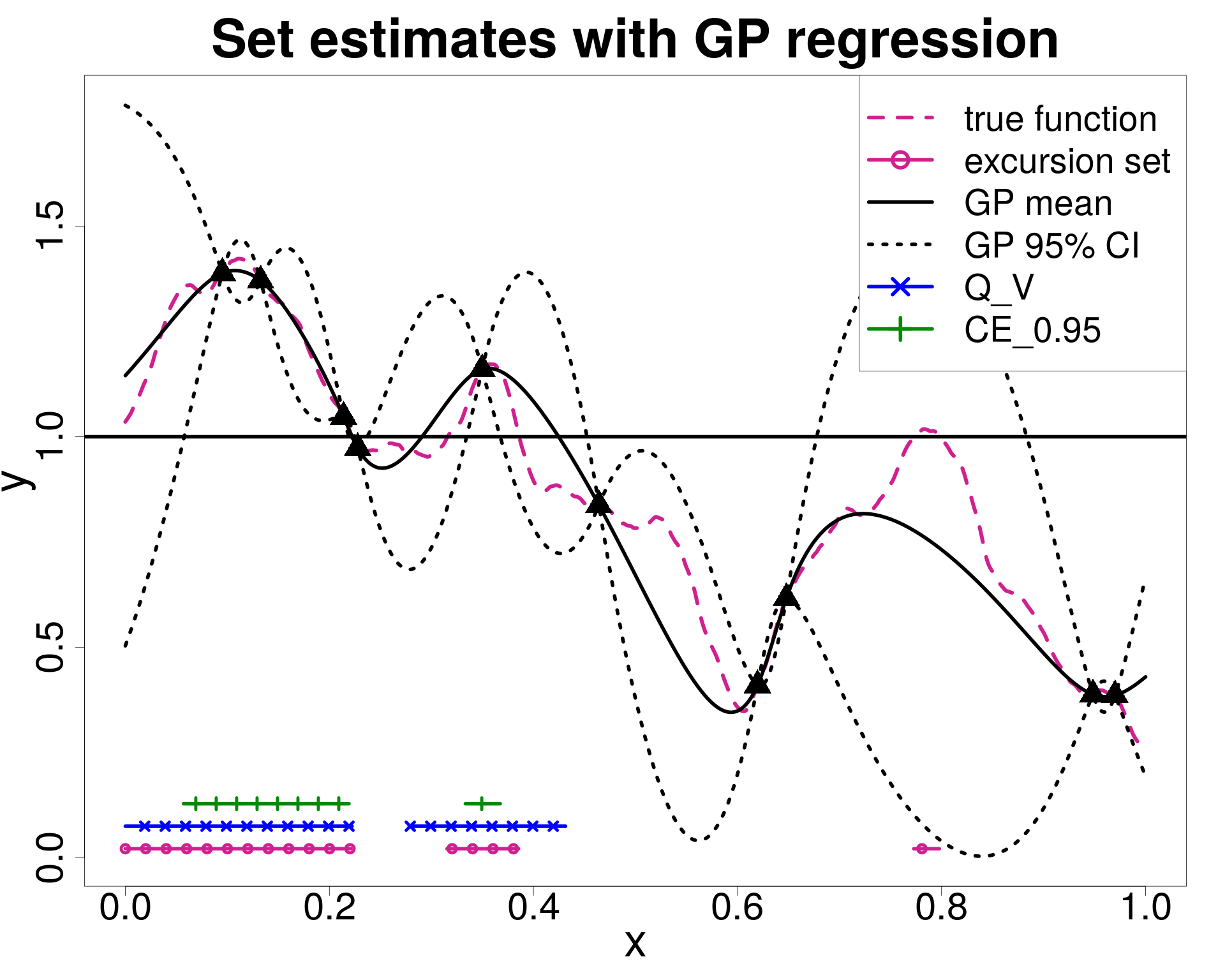}
				\subcaption{Initial DoE: maximin LHS, $n=10$.}
				\label{fig:1dExInitDoE}
			\end{minipage}\hfill\hspace{0.05cm}
			\begin{minipage}{0.48\textwidth}
				\centering
				\includegraphics[width=\linewidth]{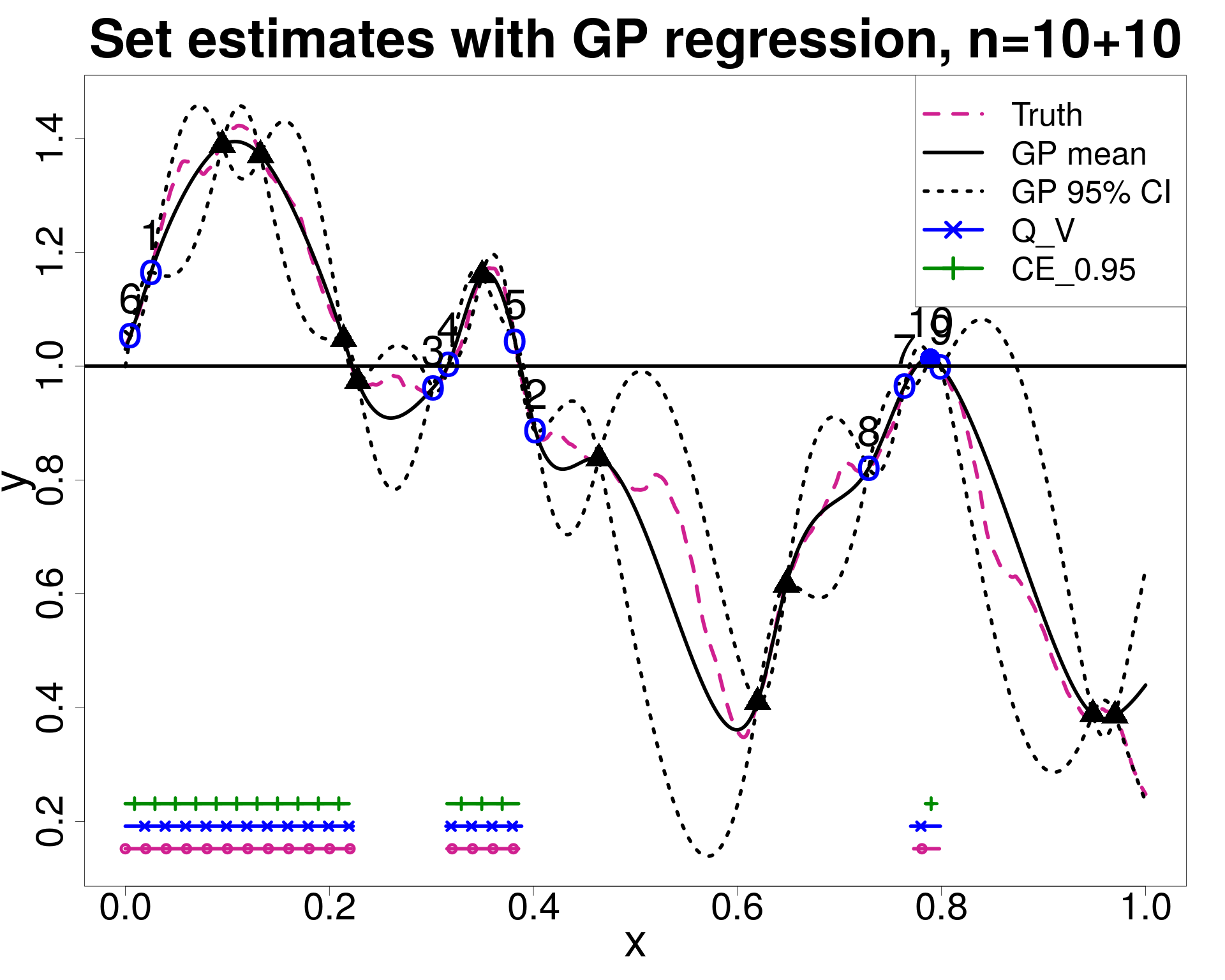}
				\subcaption{Adaptive DoE: strategy $T2$, $10$ new points.}
				\label{fig:1dExDoE10}	
			\end{minipage}
			\caption{Example of excursion set estimation. Comparison of Vorob'ev expectation ($Q_V$) and conservative estimate ($\CE_{0.95}$). The numbers near the new points indicate the order in which they were added to the DoE.}
			\label{fig:1dExample}
		\end{figure}

Consider now a generic estimate $\widetilde{\Gamma}$ for $\setOfInt$ and denote with $\volumeOf[A]$ the volume of $A \subset \inSpace$. Simple quality indicators for a set estimate are the volume of false positives $\volumeOf[\widetilde{\Gamma} \setminus \setOfInt]$, i.e.\ the volume of points estimated in the set while actually outside $\setOfInt$, and the volume of false negatives $\volumeOf[\setOfInt \setminus \widetilde{\Gamma}]$, i.e.\ the volume of points estimated not in the excursion set while actually inside. For example, in \citet{Chevalier2013} and~\citet{Chevalier.etal2014}, $\setOfInt$ is estimated with the Vorob'ev expectation, a notion borrowed from random set theory~\citep[][Chapter~2]{Molchanov2005}, that aims to minimize the overall volume of misclassified points. \Cref{fig:1dExInitDoE} shows an analytical example where the input space is $\inSpace=[0,1]$ and the function~$f$ is generated as a realization of a GP (purple dashed line) with mean zero and Mat\'ern covariance kernel with hyper-parameters $\nu=3/2$, $l=0.3$, $\sigma^2=0.3$, see, e.g.,~\citet{rasmussen2006gaussian}, Chapter~4, for details on the parametrization. We build a GP model  (black solid line) from $n=10$ evaluations of $f$ (black triangles) chosen with a maximin Latin hypercube sample (LHS) design  
and we estimate $\setOfInt$ (purple dotted horizontal line), where $T = [1,+\infty)$, with the Vorob'ev expectation ($Q_V$, middle horizontal blue line); a comparison of $Q_V$ with the true excursion set shows that here $Q_V$ has volumes of false positive ($0.084$) and negatives ($0.025$) of the same order of magnitude.

%
%

The Vorob'ev expectation, explained in more details in~\cref{subsec:VorobIntro}, gives a similar importance to false positives and false negatives. However, in a number of applications, the cost of misclassification is not symmetric with higher penalties for false positives, for instance, than for false negatives. Practitioners may be interested in set estimates which would \textit{very likely} be included in an excursion set of the form $\setOfInt = \{ x \in \inSpace : f(x) \geq t  \}$. Such a property naturally gives more importance to the minimization of (the volume of) false positives than of false negatives. 
\citet{frenchSain2013spatio} and \citet{BolinLindgren2014} introduced the concept of conservative estimates which select sets that are deliberately smaller -- in volume -- than $\setOfInt$ and are included in the excursion set with a \textit{large probability} $\alpha$. The empty set trivially satisfies this probabilistic inclusion property, therefore conservative estimates are selected as sets with maximal volume in a family of possible estimates.
A conservative estimate at level $\alpha \approx 1$ thus enforces a low probability of false positives. 

In a reliability engineering framework, the excursion set can be the set of safe configurations and a conservative estimate aims at selecting a region which is included in the safe set.
\Cref{fig:1dExInitDoE} shows a conservative estimate at level $\alpha=0.95$ ($\CE_{0.95}$, green top horizontal line). In this example, $\CE_{0.95}$ has a false positive volume equal to zero, however a much higher volume of false negative ($0.121$) than the Vorob'ev expectation. For a fixed threshold~$t$, the excursion set above~$t$ is trivially the complement of the sojourn set below~$t$. Note, however, that this does not hold for their respective set estimates. In particular, the conservative estimate of an excursion set is not the complement of the conservative estimate of the corresponding sojourn set due to the probabilistic inclusion property.

\Citet{frenchSain2013spatio} and \citet{BolinLindgren2014} proposed an approach to compute conservative estimates for a fixed Design of Experiments (DoE). However, to the best of our knowledge, there is no study on how to reduce the uncertainty on conservative estimates with adaptive strategies. Here we focus on the problem of sequentially choosing evaluation points in order to reduce the uncertainty on conservative estimates. In order to illustrate this concept, consider the example introduced in~\cref{fig:1dExInitDoE} and notice how conservative set estimates do not intercept the excursion near $x=0.8$. In this case we can increase the size of our design of experiments by adaptively choosing new evaluations of $f$ that help us better localize the excursion. \Cref{fig:1dExDoE10} shows an example of such adaptive DoEs where, starting from the initial DoE in~\cref{fig:1dExInitDoE}, $10$ additional points are selected with strategy $C$ introduced in~\cref{sec:SequentialStrategies}.
 
 Previous adaptive DoE strategies for excursion set estimation mainly focused on recovering the boundaries of the set. In particular, \citet{Picheny2010} introduced the targeted $\IMSE$ (integrated mean squared error), $\tIMSE$, criterion to add points at locations that improve the accuracy of the model around a certain level of the response variable. \citet{Bect.etal2012} investigated the concept of Stepwise Uncertainty Reduction (SUR) strategies for GP~\citep[see also][]{vazquez2009sequential,Chevalier.etal2014,Bect.etal2017}. Such strategies, however, do not provide any control on false positives and as such are not adapted to the conservative estimation case. Here, by shifting the focus on the control of false positives, we extend the conservative estimation framework introduced by~\citet{BolinLindgren2014} to sequential design of experiments. For example, notice how in \cref{fig:1dExDoE10}, some points (e.g. numbers $1,2,8$) are chosen far from the boundary, 
 in order to improve the confidence on the classification of those regions. Here we consider a definition of conservative estimates well suited to excursion sets of Gaussian processes and we provide a SUR strategy with tractable criteria to reduce the uncertainty on conservative estimates. The adaptive strategies are introduced in the case of excursion sets above~$t\in \R$, however our $\RprogLang$ implementation, available on CRAN, allows also for excursions below~$t$.

\subsection{Outline of the paper}

The remainder of the paper is structured as follows. In the next section we briefly recall some background material. In particular,~\cref{subsec:VorobIntro} reviews set estimates preliminary to this work and~\cref{subsec:SoaSequentialGP} recalls the concept of SUR strategies. 
In~\cref{subsec:UQconsEst} we introduce the metrics used to quantify uncertainty on such estimates. In~\cref{subsec:surCriteria}, we detail the proposed sequential strategies and, in~\cref{sec:SeqProofs}, we derive fast-to-compute formulae for the associated criteria and illustrate their implementation. \Cref{sec:TestCase} presents a benchmark study where we analyze a trade-off between noisy evaluations and batch size in three scenarios.  \Cref{sec:IRSNtestCase} shows the results obtained on a reliability engineering test case. In \cref{sec:PreliminariesAddRes} we provide more properties for conservative estimates that further justify the choices made in \cref{subsec:VorobIntro}. In supplementary material we also apply the proposed strategies on a coastal flood problem. All proofs are in~\cref{sec:PreliminariesAddRes,sec:SeqProofs}.

\section{Background}
\label{sec:Background}

Let us consider $n$ observations of the function~$f$, possibly tampered by measurement noise
\begin{equation}
	z_i = f(x_i)+ \tau \epsilon_i \qquad x_i \in \inSpace, \quad i =1, \ldots, n
	\label{eq:obsModel}
\end{equation}
with $\epsilon_i$ independent realizations of standard Gaussian measurement noise and $\tau^2$ a known homogeneous noise variance. 

In a Bayesian framework ~\citep[see, e.g.,][and references therein]{Chiles.Delfiner2012} we consider $f$ as a realization of an almost surely continuous Gaussian process (GP) $\xi \sim GP(\mean,\covkern)$, with mean function~$\mean(x) := \E[\xi_x]$ and covariance function~$\covkern(x,x^\prime) := \operatorname{Cov}(\xi_x,\xi_{x^\prime}), \ x,x^\prime \in \inSpace$. The mean function could potentially have a structure such as $\mean(x) = \sum_{i=1}^p \beta_j f_j(x)$, where $\boldsymbol{\beta} = (\beta_1, \ldots, \beta_p) \in \R^p$ are parameters to be estimated and $f_j$ are known basis functions. With this notation, $z_i$ is a realization of $Z_i = \xi_{x_i} + \tau\varepsilon$ where $\varepsilon \sim N(0,1)$.  
For $\nNot>0$, we denote by $\mathbf{z}_{\nNot} = (z_1, \ldots, z_\nNot) \in \outSpace^n$ the observations at an initial design of experiments (DoE) $\doe_\nNot = (x_1, \ldots, x_\nNot) \in \inSpace^\nNot$. The posterior distribution of the process is Gaussian with mean and covariance computed as the conditional mean $\mean_n$ and conditional covariance $\covkern_n$ given the observations, see, e.g.,~\citet{Santner.etal2018}, Chapter~4, for closed-form formulae.	

\subsection{Vorob'ev expectation and conservative estimates}
\label{subsec:VorobIntro}

The prior distribution on $\xi$ induces a (random) set $\Gamma(\xi) =\{ x \in \inSpace : \xi_x \in T \}$. We will omit the dependency on $\xi$ when obvious and refer to $\Gamma(\xi)$ as $\Gamma$ when appropriate. By using the posterior distribution of $\xi$, we can provide estimates for $\setOfInt$. See, e.g.~\citet{Chevalier.etal2014}, \citet{BolinLindgren2014} and \citet{Azzimonti2016} for summaries of different approaches.
A central tool for the approach presented here is the \textit{coverage probability function} of a random closed set $\Gamma$, defined as
\begin{equation*}
p(x) = P(x \in \Gamma), \ x \in \inSpace.
\end{equation*}
In our case we consider the posterior coverage function~$p_{\nNot}$, defined with the posterior probability $P_\nNot(\cdot) = P(\cdot \mid \mathbf{Z}_{\nNot}=\mathbf{z}_{\nNot})$, where $\mathbf{Z}_\nNot = (Z_1, \ldots, Z_\nNot)$. If $T=(-\infty,t]$, then $p_{n}(x)= \Phi\left(\frac{t-\mean_n(x)}{\sd_n(x)}\right)$, where $\Phi(\cdot)$ is the CDF of a standard Normal random variable and $\sd_n(x) =\sqrt{\covkern_n(x,x)}$. 
The coverage function defines the family of \emph{Vorob'ev quantiles}
\begin{equation}
Q_{\nNot,\rho} = \{x \in \inSpace: p_{\nNot}(x)\geq \rho\},
\label{eq:vorobQuantiles}
\end{equation}
with $\rho \in [0,1]$. These sets are closed for each $\rho \in [0,1]$ \citep[see][Proposition~1.34]{Molchanov2005} and form a family of possible estimates parametrized by $\rho$. 

The level $\rho$ can be selected in different ways. The choice $\rho = 0.5$ leads to the \emph{Vorob'ev median}, which is not conservative. \emph{Vorob'ev expectation}~\citep{Vorobev84,Molchanov2005,Chevalier.etal2013b} relies on the notion of measure. In the example in~\cref{fig:1dExDoE10} and in the applications presented here we use the standard volume, however here we introduce the concept in a slightly more general form by using a finite measure $\measure$, for example, $\measure$ could be a probability distribution on $\inSpace$. The Vorob'ev expectation is defined as the quantile $Q_{\nNot,\rho_V}$ such that $\measureOf[Q_{\nNot,\rho}] \leq \E[\measureOf[\Gamma]] \leq \measureOf[Q_{\nNot,\rho_V}]$ for all $\rho > \rho_V$. 
The set $Q_{\nNot,\rho_V}$ is also the minimizer of $\E[\measureOf[\Gamma \Delta M]]$\footnote{For any set $A,B$, $A \Delta B := (A\setminus B)\cup (B\setminus A)$} among all measurable sets such that $\measureOf[M] = \E[\measureOf[\Gamma]]$,  
see, e.g., \citet[][Theorem~2.3, Chapter~2]{Molchanov2005}. Vorob'ev expectation minimizes a uniformly weighted combination of the expected measure of false positives ($\E[\measureOf[M\setminus \Gamma]]$, also called \emph{type I error}) and false negatives ($\E[\measureOf[\Gamma \setminus M]]$, \emph{type II error}) among sets with measure equal to $\E[\measureOf[\Gamma]]$. In~\cref{subsec:CEvorobevQuant} we prove a similar result for generic Vorob'ev quantiles. The quantity $\E[\measureOf[\Gamma_1\Delta \Gamma_2]]$, for two random sets $\Gamma_1,\Gamma_2 \subset \inSpace$, is often called \emph{expected distance in measure}. \citet{Chevalier2013} used this distance to adaptively reduce the uncertainty on Vorob'ev expectations. In~\cref{subsec:UQconsEst} we adapt it for conservative estimates.

\begin{figure}
	\centering
	\begin{minipage}{0.5\textwidth}
		\centering
		\includegraphics[width=\linewidth]{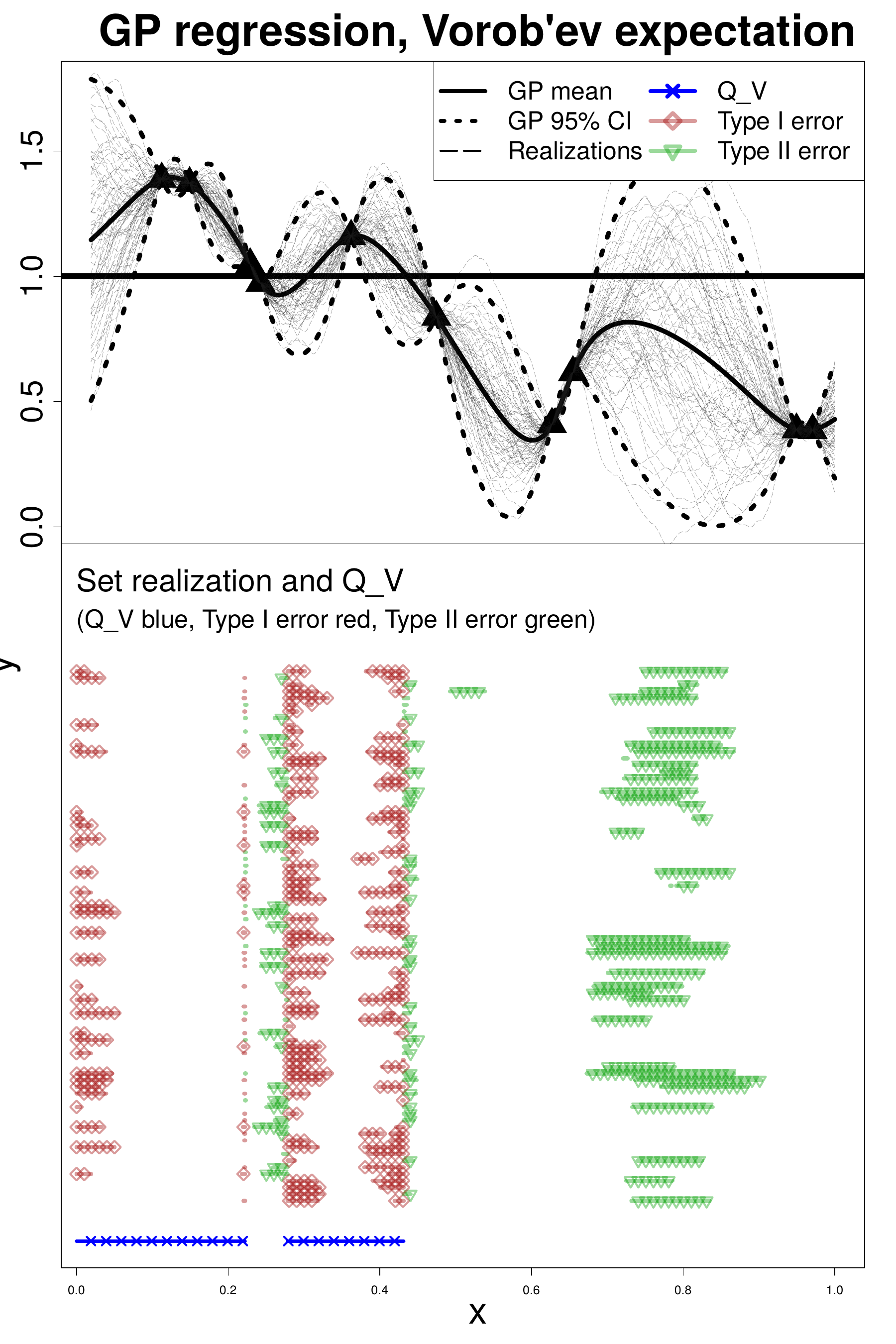}
		\subcaption{Vorob'ev expectation ($Q_V$, blue dotted).}
		\label{fig:1dEx_1}
	\end{minipage}\hfill 
	\begin{minipage}{0.5\textwidth}
		\centering
		\includegraphics[width=\linewidth]{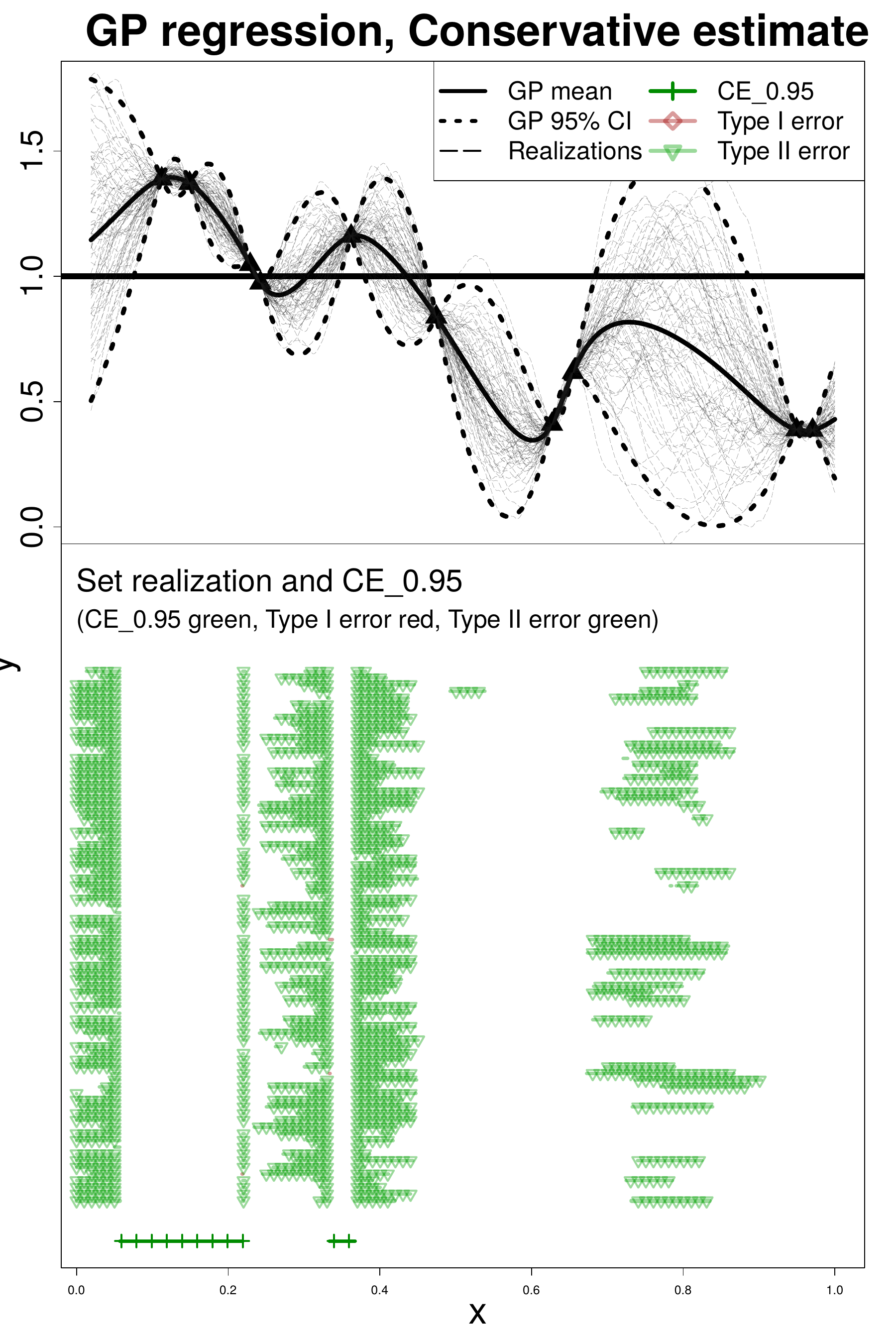}
		\subcaption{Conservative estimate ($\alpha=0.95$, green).}
		\label{fig:1dEx_2}	
	\end{minipage} 
	\caption{1-dimensional example, $n=10$ evaluations. Type I (red, diamonds) and Type II (green, triangles) errors for Vorob'ev expectation and conservative estimate.} 
	\label{fig:1dEx}
\end{figure} 

\begin{table}
	\caption{\label{tab:1dExample}Summary values for example in~\cref{fig:1dEx}, estimated from 100 GP realizations.} 
	\centering
	\fbox{
		\begin{tabular}{ c c  c  c c }
			& $\rho$ & Type I error (mean $\pm$ sd)  & Type II error (mean $\pm$ sd) & $\hat{P}(Q_\rho \subset \Gamma)$ \\
			\hline 
			$Q_{\rho_V}$ & 0.393 & $0.046 \pm 0.029$   & $0.053 \pm 0.058$ & $0.02$ \\[2pt]
			$Q_{0.5}$ & 0.500	& $0.035 \pm 0.026$ & $0.061 \pm 0.058$ & $0.07$ \\[2pt]
			$Q_{0.95}$ & 0.950 & $5.7\times10^{-4} \pm 1.9\times10^{-3}$ & $0.168 \pm 0.063$ & $0.87$ \\[2pt]
			$\CE_{0.95}$ & 0.987 & $9.5\times10^{-5} \pm 4.7\times10^{-4}$ & $0.187 \pm 0.063$  & $0.95$ \\[2pt]
		\end{tabular}
	}
\end{table}

%
%

Conservative estimates~\citep{BolinLindgren2014,frenchSain2013spatio} embed probabilistic control on false positives in the estimator. Denote with $\mathfrak{C}$ a family of closed subsets in $\inSpace$. A \emph{conservative estimate at level $\alpha$} for $\setOfInt$ is a set $\CE_{\alpha,\nNot}$ defined as  
\begin{equation}
\CE_{\alpha,\nNot} \in \underset{C \in \mathfrak{C}}{\arg\max} \{ \measureOf[C]  : P_{\nNot}(C \subset \Gamma) \geq \alpha\}.
\label{eq:genericConservative}
\end{equation}
The set $\CE_{\alpha,\nNot}$ is therefore a maximal set (according to $\measure$) in the family $\mathfrak{C}$ such that the posterior probability of inclusion is at least $\alpha$. Here, by following~\citet{frenchSain2013spatio}, \citet{BolinLindgren2014} and \citet{Azzimonti.etal2016b}, we choose $\mathfrak{C}$ as the family of Vorob'ev quantiles $\{Q_{\nNot,\rho} : \rho \in [0,1] \}$ as introduced in~\cref{eq:vorobQuantiles}. While the concept of probabilistic inclusion might seem unusual at first, conservative estimates are actually linked with the well known concept of confidence regions, as we briefly show in~\cref{subsec:confidenceRegs}. Note further that the condition $P_n(C \subset \Gamma) = P_n(C \setminus \Gamma = \emptyset) \geq \alpha$ controls the probability of false positives. We can visualize this property on the one dimensional example introduced in \cref{fig:1dExample} by empirically estimating the expected measure of false positives, $\E_n[\measureOf[\CE_{0.95} \setminus \Gamma]]$. \Cref{fig:1dEx} shows $80$ posterior realizations of the GP (light dashed black lines) and for each realization we computed the false positive (type I error, horizontal red lines) and false negatives (type II error, horizontal green lines). Notice how they are symmetrically minimized by the Vorob'ev expectation (\cref{fig:1dEx_1}) while the conservative estimate with $\alpha=0.95$ (\cref{fig:1dEx_2}) has small false positives and much larger false negatives. \Cref{tab:1dExample} reports the values for the expected volume of type I and II errors and the estimated probability of inclusion, $\hat{P}(Q_\rho \subset \Gamma)$. The Vorob'ev expectation may be closer to the truth than conservative estimates, especially for small DoEs, however $\CE_{\alpha, \nNot}$ gives control on the probability of false positives. \Cref{tab:1dExample} also reports the values for the Vorob'ev quantile $Q_{0.95}$, i.e.\ a non adaptive high quantile choice for $\rho$. Note that $P(Q_{0.95} \subset \varGamma) < 0.95$, in fact, the quantile's definition based on the marginal probability $p_n(x) \geq 0.95$, $x \in \inSpace$, does not imply any statement on the joint probability of inclusion. 

The computation of $\CE_{\alpha,n}$ in~\cref{eq:genericConservative} requires finding a set $C$ of maximum measure among sets included in the random set $\Gamma$ with probability at least $\alpha$. 
When $\mathfrak{C}$ is the family of Vorob'ev quantiles $Q_\rho$, $\rho \in [0,1]$, this optimization can be solved with a simple dichotomic search on $\rho$. See \cref{subsec:CEvorobevQuant} for more details. If $T=(-\infty,t]$, for $\rho\in [0,1]$, we approximate $P_n(Q_\rho \subset \Gamma) \approx P_n(\xi_{q_1} \leq t, \ldots, \xi_{q_\ell} \leq t)$, where $\{q_1, \ldots, q_\ell \} \subset Q_\rho$  is a set of $\ell$ points in $Q_\rho$ with $\ell$ large. The probability above is then computed efficiently with the integration technique proposed by~\citet{Azzimonti.etal2016b}. The  number $\ell$ is generally chosen as large as the computational budget allows. This technique can also be used for excursion sets above~$t$. An alternative method, not used here, is Monte Carlo with conditional realizations of the field, see, e.g.~\citet{Azzimonti.etal2016} for fast approximations of conditional realizations. 
The optimal Vorob'ev level chosen for conservative estimates at level $\alpha$ is denoted by $\ConsLevel$ in what follows : $\CE_{\alpha,n} := Q_{n,\ConsLevel}$.


%


\subsection{SUR strategies}
\label{subsec:SoaSequentialGP}

Sequential design of experiments adaptively chooses the next evaluation points according to a strategy with the aim of improving the estimation of a quantity (or set, here) of interest. As shown in the introduction we can improve the set estimates in~\cref{fig:1dExInitDoE} by carefully adding new function evaluations, see~\cref{fig:1dExDoE10}. There are many ways to build a sequential DoE, see, e.g.,~\citet{Santner.etal2018}, chapter~6. Here we follow the Stepwise Uncertainty Reduction approach~\citep[SUR, see, e.g.,][]{fleuret1999graded,Bect.etal2012,Chevalier.etal2014,bect.etal2016} and select a sequence of points in order to reduce the uncertainty of a quantity of interest. 
%

In the remainder of the paper we consider that the first $\nNot$ points $x_1, \ldots, x_n$ and the respective evaluations $\mathbf{z}_\nNot$ are known and we denote by $\E_n[\cdot]= \E[\cdot \mid \mathbf{Z}_\nNot = \mathbf{z}_{\nNot}]$ the expectation conditional on $\mathbf{Z}_\nNot=\mathbf{z}_{\nNot}$. 
We are interested in selecting the next batch of $q$ locations $x_{n+1}, \dots, x_{n+q}$. The advantage of batches with $q>1$ lies in the fact that parallel function evaluations, when available, can save the user wall-clock time.  In a sequential setting the response values at these points are unknown before evaluations, therefore we denote by 
$\E_{n,\newEval[q]}[\cdot]$ the conditional expectation given the first $n$ evaluations and with the next locations fixed at $\newEval[q] = (x_{n+1}, \ldots, x_{n+q}) \in \inSpace^q$. 

For a specific problem,  we consider a quantity, denoted by $H_n$, which measures the residual uncertainty at step $n$. We define this quantity for the conservative estimation problem in~\cref{subsec:UQconsEst}. If the first $n$ locations and evaluations are known, then $H_n$ is a (deterministic) real number quantifying the residual uncertainty on the estimate. As an example, consider the setup in~\cref{fig:1dExample} and the uncertainty $H_n := \E_n[\measureOf[\Gamma \Delta \CE_{0.95}]]$; we can compute $H_n$, $n=10$, with numerical integration and obtain $H_{10}=0.23$. On the other hand, the quantity $H_{n+1}$, seen from step $n$, is random because $Z_{n+1}$ is random. 
%
The next batch of $q$ locations can then be selected following the principles of a SUR strategy, i.e.\ by setting
\begin{equation}
\mathbf{x}^*_{n+q} \in \arg \min_{\newEval[q] \in \inSpace^q} \E_{n,\newEval[q]}[H_{n+q}],
\label{eq:surGeneric}
\end{equation}
a minimizer of the future uncertainty in expectation. For a more complete and theoretical perspective on SUR strategies see, e.g.,~\citet{bect.etal2016} and references therein.
There are many ways to proceed with the minimization introduced above. See, e.g., \citet{osborne2009gaussian}, \citet{Ginsbourger2010towards}, \citet{Bect.etal2012}, \citet{gonzalez2016glasses} and references therein. 
The objective function in~\cref{eq:surGeneric} is a \emph{batch sequential one-step lookahead sampling criterion} and is denoted by $J_n : \newEval[q] \in \inSpace^q \mapsto \E_{n,\newEval[q]}\left[H_{n+q} \right] \in \R$. 


%

\begin{figure}
	\begin{minipage}{0.48\textwidth}
		\centering
		\includegraphics[width=\linewidth]{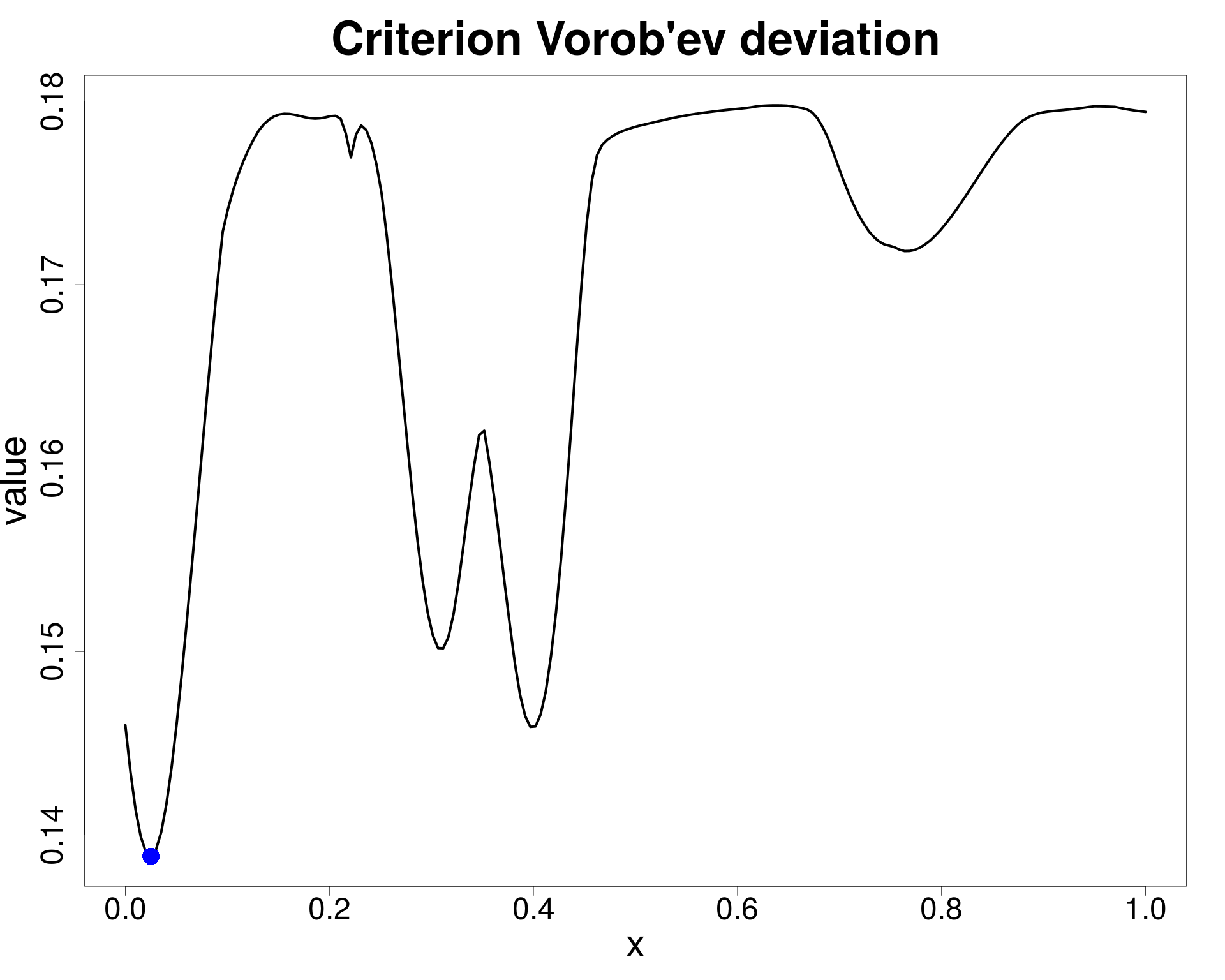}
		\subcaption{Criterion $J_n$. The next point is the minimizer of this function, blue dot.}
		\label{fig:1dExT2criterion}
	\end{minipage}\hfill\hspace{0.05cm}
	\begin{minipage}{0.48\textwidth}
		\centering
		\includegraphics[width=\linewidth]{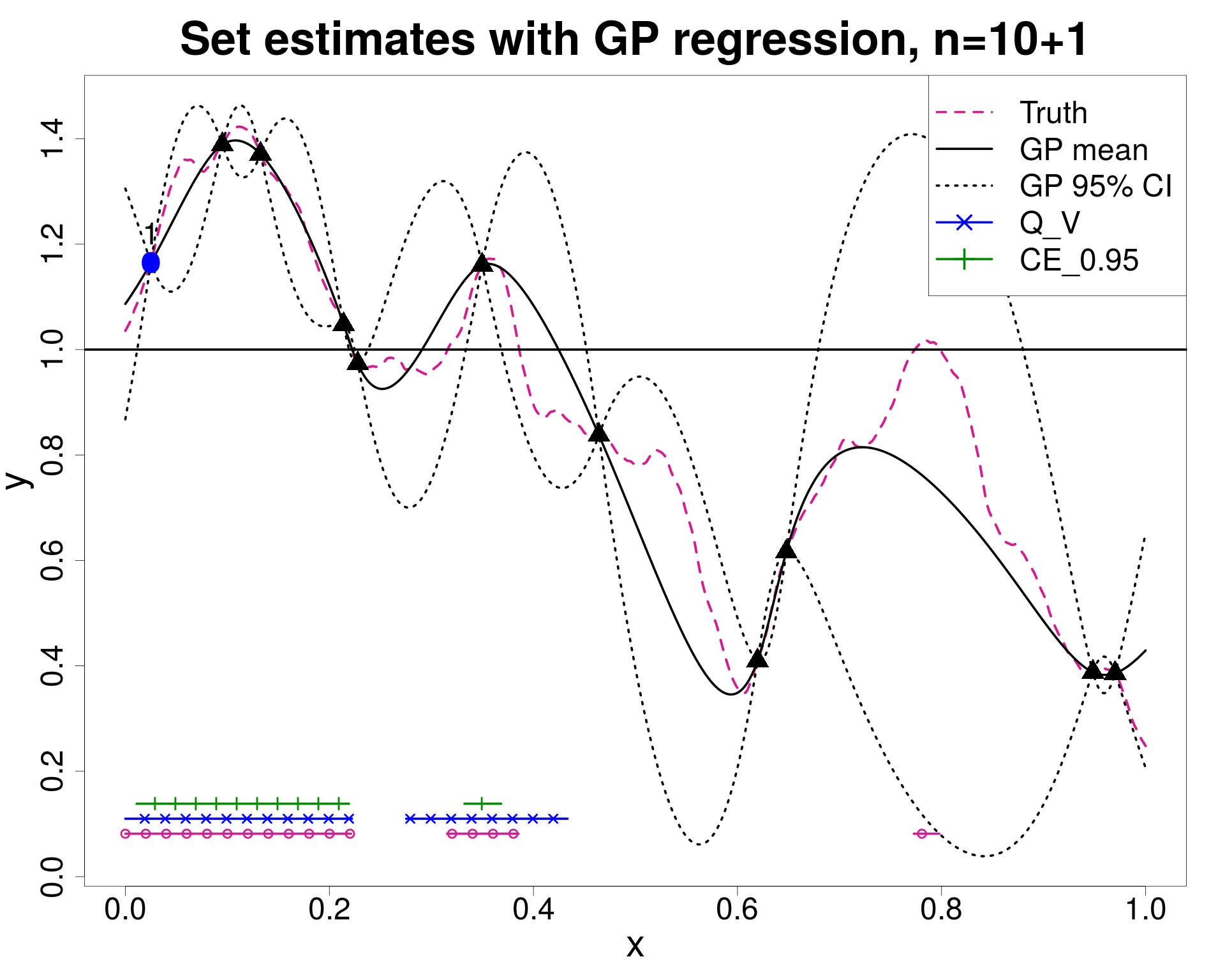}
		\subcaption{Adaptive DoE with $n=11$, the last point (blue dot) is chosen with $J_n$.}
		\label{fig:1dExDoE11}	
	\end{minipage}
	\caption{Adaptive DoE with SUR strategy on the example introduced in~\cref{fig:1dExample}.}
	\label{fig:1dExSUR}
\end{figure} 
We can build a SUR strategy with the uncertainty $H_n= \E_n[\measureOf[\Gamma \Delta \CE_{0.95}]]$ mentioned above. The criterion associated with this uncertainty has the remarkable property that it can be computed with fast-to-evaluate formulae, thus making its optimization more convenient. \Cref{fig:1dExT2criterion} shows the function~$J_n$ for $n=10$ and $q=1$; the next evaluation $x_{11}$ is chosen as the minimizer of this function restricted to a finite discretization of the domain. \Cref{fig:1dExDoE11} shows the updated GP model and $\CE_{0.95}$ which, compared to \cref{fig:1dExInitDoE}, is now larger while still included inside the true set.

The expectation $\E_n$ can only be computed if we know $\covkern$ which is often chosen from a parametric family depending on few hyper-parameters, $l$ and $\sigma$ in the analytical example. In practice, the hyper-parameters are unknown and can be estimated with a plug-in or with a fully Bayesian approach. In this work we follow the previous literature on boundary estimation with GP models \citep[see, e.g.][]{ranjan2008sequential,picheny2013quantile,Chevalier.etal2014,Azzimonti.etal2016b} and we use plug-in maximum likelihood estimates for the hyper-parameters. Model checking procedures, such as cross-validation, can be used to evaluate the robustness of hyper-parameters' estimates. If only few observations are available, a fully Bayesian approach might better capture the overall uncertainty. However such an approach is not straightforward for many SUR criteria \citep[see, e.g.,][]{Stroh2018} and it involves an additional computational cost. 


In the next section we detail two uncertainty functions tailored for conservative estimates and we show how their respective SUR criteria can be computed. 




\section{SUR strategies for conservative estimates}
\label{sec:SequentialStrategies}

\subsection{Uncertainty quantification on conservative estimates}
\label{subsec:UQconsEst}
	
	Our object of interest is $\setOfInt$, therefore we require uncertainty functions that take into account the whole set. \citet{Chevalier.etal2013b}~and~\citet{Chevalier2013} evaluate the uncertainty on the Vorob'ev expectation with the \emph{Vorob'ev deviation}, i.e.\ the expected distance in measure between the current estimate $Q_{\nNot,\rho_\nNot}$ and the set $\Gamma$. In this section we introduce an uncertainty suited for conservative estimates. The idea is to describe the uncertainty by looking at the expected measure of false negatives. In the example of~\cref{fig:1dEx_2}, this quantity is the mean measure of the sets in green. Expected distance in measure and false negatives are related concepts and, in order to highlight this connection, let us first recall \citep{Chevalier2013} that the Vorob'ev deviation of a quantile $Q_{\nNot,\rho}$ is 
		\begin{equation}
		H_{\nNot,\rho} = \E_\nNot[\measureOf[\Gamma \Delta Q_{\nNot,\rho}]] = \E_\nNot[\measureOf[Q_{\nNot,\rho} \setminus \Gamma] ] + \E_\nNot[\measureOf[\Gamma \setminus Q_{\nNot,\rho}] ], \qquad \rho \in [0,1].
		\label{eq:VorobUncertainty}
		\end{equation}

	In the following sections, $\rho$ will be chosen either as the Vorob'ev median, $\rho=0.5$, or as the threshold for a conservative estimate at level $\alpha$ after $n$ evaluations, $\rho = \ConsLevel_{\nNot}$.

	
	
%
	Let us denote by $G^{(1)}_\nNot(\rho) = \measureOf[Q_{\nNot,\rho} \setminus \Gamma]$ and $G^{(2)}_\nNot (\rho)= \measureOf[\Gamma \setminus Q_{\nNot,\rho}]$ the random variables associated with the measure of the first and the second set difference in~\cref{eq:VorobUncertainty} and recall that their expectations, i.e.\ $\E_\nNot[ G^{(1)}_\nNot(\ConsLevel_\nNot) ]$ and $\E_\nNot[G^{(2)}_\nNot(\ConsLevel_\nNot)]$, are called \emph{Type I} and \emph{Type II} errors respectively. 
	Type II error provides a quantification of the residual uncertainty on the conservative estimate; we formalize this concept with the following definition.
	\begin{definition}[Type II uncertainty]
		Consider the Vorob'ev quantile $Q_{\nNot,\ConsLevel_\nNot}$ corresponding to the conservative estimate at level $\alpha$ for $\Gamma$. The \emph{Type II uncertainty} is defined as
		\begin{equation}
		H^{\textsc{t2}}_{\nNot,\ConsLevel_\nNot} := \E_\nNot[G^{(2)}_\nNot(\ConsLevel_\nNot)] = \E_\nNot[\measureOf[\Gamma \setminus Q_{\nNot,\ConsLevel_\nNot}]].
		\label{eq:T2Uncertainty}
		\end{equation}
		\label{def:eqT2Uncertainty}
	\end{definition}
	
	This definition of residual uncertainty is reasonable for conservative estimates because they aim at controlling the error $\E_\nNot[ G^{(1)}_\nNot(\ConsLevel_\nNot) ]$. In particular it is possible to show that the ratio between the Type I error and the measure of a conservative estimate is bounded by a constant which is close to zero when $\alpha$ is close to one. 
	
	\begin{proposition}
		Consider the conservative estimate $Q_{\nNot,\ConsLevel_\nNot}$, then the ratio between the error $\E_\nNot[ G^{(1)}_\nNot(\ConsLevel_\nNot) ]$ and the measure $\measureOf[Q_{\nNot,\ConsLevel_\nNot}]$ is bounded by $1-\alpha$.
		\begin{proof}
			See~\cref{sec:PreliminariesAddRes}.
		\end{proof}
		\label{prop:boundT1error}
	\end{proposition}

	\begin{figure}
		\begin{minipage}{0.48\textwidth}
			\centering
			\includegraphics[width=\linewidth]{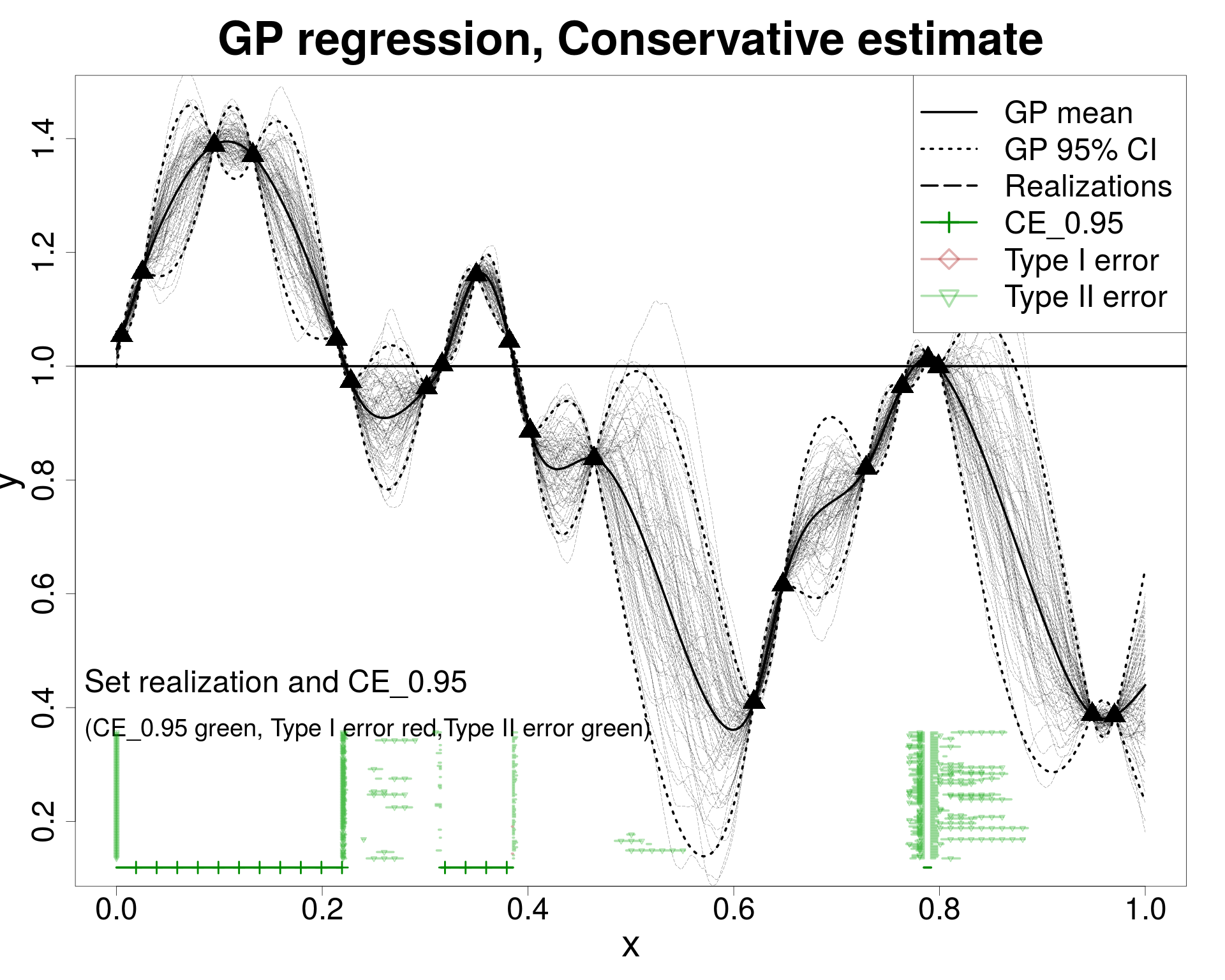}
			\subcaption{Type I/II errors final model.}
			\label{fig:1dExT2Uncertainty}
		\end{minipage}\hfill\hspace{0.05cm}
		\begin{minipage}{0.48\textwidth}
			\centering
			\includegraphics[width=\linewidth]{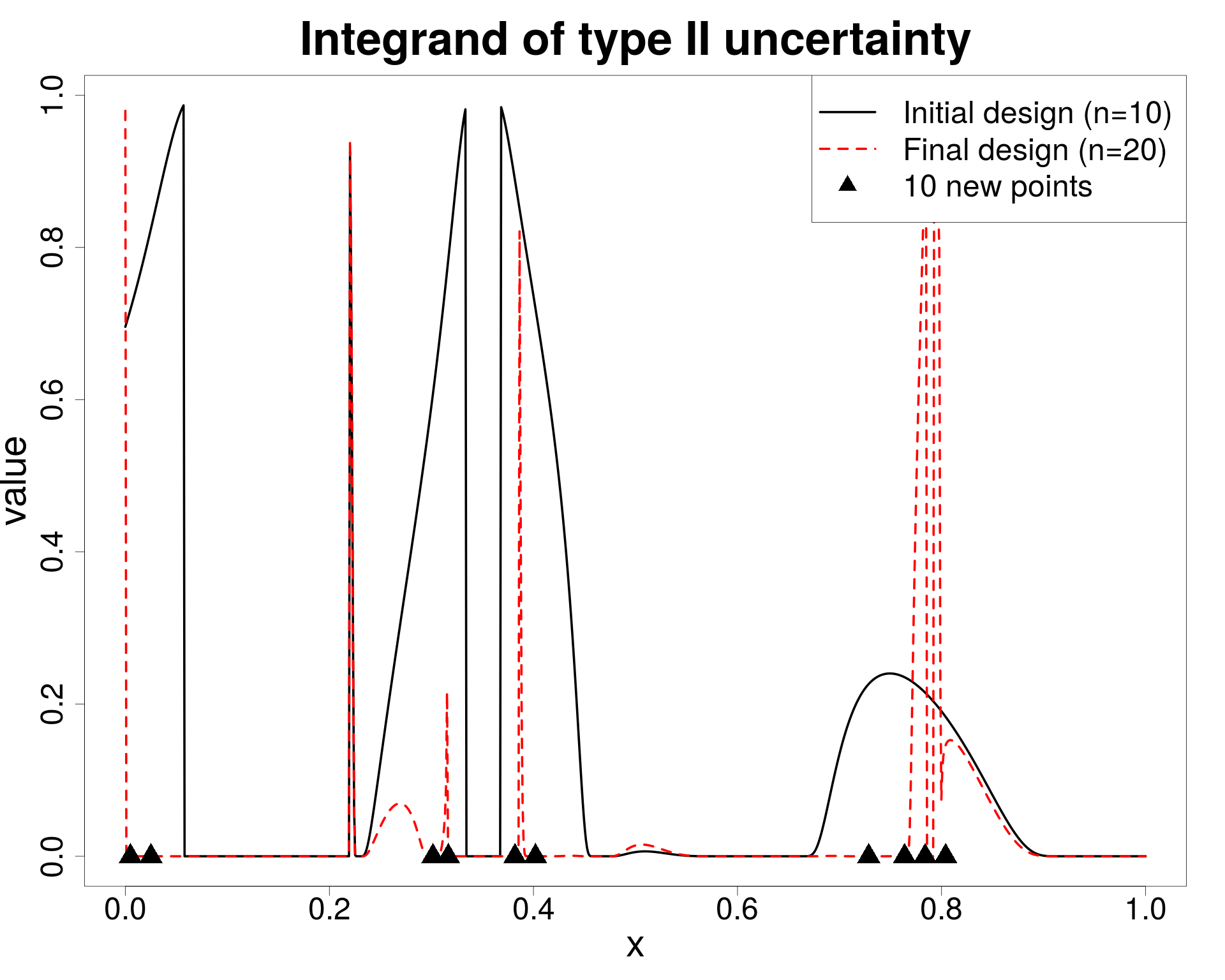}
			\subcaption{Integrand of type II uncertainty.}
			\label{fig:1dExT2comparison}	
		\end{minipage}
		\caption{1d example, DoE with initial points of \cref{fig:1dExDoE10} plus $10$ adaptive new points.}
		\label{fig:1dExT2addendum}
	\end{figure} 

	If the posterior GP mean provides a good approximation of the function~$f$, conservative estimates with high $\alpha$ tend to be inside the true set $\setOfInt$. In such situations the Type I error is usually very small 
	while Type II error could be rather large. Note the differences in Type I/II errors reported in \cref{tab:1dExample} for the analytical example. 
	Type II uncertainty is thus a relevant quantity when evaluating conservative estimates. In the test case studies we also compute the expected type I error to check that it is consistently small.
	
	\Cref{fig:1dEx_2} provides a visualization of type II uncertainty: the green horizontal lines are realizations of $\Gamma \setminus Q_{\nNot,\ConsLevel_\nNot}$  obtained from posterior GP draws. Type~II uncertainty is the expected value of the measure of such sets. In the example shown, this amounts to  $0.187$.  
	Consider now the updated GP estimate in \cref{fig:1dExDoE10} where $10$ new points were added to the initial DoE by following an adaptive strategy that will be described later. \Cref{fig:1dExT2Uncertainty} shows how the type II uncertainty is reduced in the updated model: the green horizontal lines are much shorter, resulting in a smaller expected measure of  $0.035 \pm 0.022$. 
	
	The expectation and integration operators can be exchanged in~\cref{eq:T2Uncertainty}, therefore type II uncertainty can be further written as 
	\begin{equation}
	H^{\textsc{t2}}_{\nNot,\ConsLevel_\nNot} = \E_\nNot[\measureOf[\Gamma \setminus Q_{\nNot,\ConsLevel_\nNot}]] = \int_{\inSpace}p_n(x) \mathds{1}_{(Q_{\nNot,\ConsLevel_\nNot})^C}(x) d\measureOf[x]
	\label{eq:T2Uncwithpn}
	\end{equation}
	where $\mathds{1}_{A^C}$ denotes the indicator function of the complement of set $A$. \Cref{fig:1dExT2comparison} plots the integrand in~\cref{eq:T2Uncwithpn} for the example in~\cref{fig:1dExDoE10} with $n=10$, after the initial LHS design, (black solid line), and for $n=20$ (red dashed line),  after $10$ points were added with the same adaptive strategy as in~\cref{fig:1dExT2Uncertainty}. The large bumps shown for $n=10$ are reduced to small spikes after the new points are added in appropriate locations.

\subsection{SUR criteria}
\label{subsec:surCriteria}

Suppose that the first $n$ locations and their respective function evaluations are known. Here we introduce one-step lookahead SUR criteria for conservative estimates based on the measures of residual uncertainty previously introduced. In a sequential algorithm we minimize such criteria to select the next batch of $q>0$ locations $x_{n+1}, \dots, x_{n+q} \in \inSpace$.

Since the locations $x_{n+1}, \ldots, x_{n+q}$ and the responses $Z_{n+1}, \ldots, Z_{n+q}$ are unknown, the uncertainty  $H_{n+q}$ and the conservative level $\ConsLevel_{n+q}$ are random variables. The criteria introduced below (\cref{eq:VorobCriterion,eq:T2criterion}) are properly defined for $\rho=\ConsLevel_{n+q}$, the conservative level after $n+q$ evaluations, however, in that case the expectations involved can only be computed with an expensive Monte Carlo procedure. The criteria's implementations use the last known level $\rho=\ConsLevel_{n}$ which allows to expand the criteria in fast-to-evaluate formulae. We consider two sampling criterion based on the uncertainty functions in~\cref{eq:VorobUncertainty,eq:T2Uncertainty}. 

The \emph{conservative $J_n$} criterion is defined as	
\begin{align}
	J_n(\newEval[q];\ConsLevel_{n}) =  \E_{n,\newEval[q]} \left[ H_{n+q, \ConsLevel_{n} } \right] 
	= \E_{n,\newEval[q]} \left[\measureOf[\Gamma \Delta Q_{n+q,\ConsLevel_{n}}] \right]
	\label{eq:VorobCriterion}
\end{align}
for $\newEval[q] = (x_{n+1}, \dots, x_{n+q}) \in \inSpace^q$, where $Q_{n+q,\ConsLevel_{n}}$ is the Vorob'ev quantile obtained with $n+q$ evaluations of the function at level $\ConsLevel_{n}$, the conservative level obtained with $n$ evaluations. This is an adaptation of the Vorob'ev criterion introduced by \citet{Chevalier2013} based on the Vorob'ev deviation \citep{Vorobev84,Molchanov2005,Chevalier.etal2013b}. In Chapter~4.2, \citet{Chevalier2013}, derives the formula for this criterion for the Vorob'ev expectation, i.e.\ the quantile at level $\rho=\rho_{n,V}$. 

Note that each evaluation of $J_n(\newEval[q])$ requires calculating the expectation $\E_{n,\newEval[q]}[\cdot]$. This could, in principle, be achieved with a Monte Carlo procedure that draws samples from the posterior distribution of $Z$, generate posterior samples for $\Gamma$ and uses such samples to empirically evaluate the expectation. This procedure however could potentially be very costly and, since many evaluations of $J_n$ are required in order to find its minimum, we provide in~\cref{sec:SeqProofs},~\cref{prop:VorobCrit} a fast-to-evaluate formula to compute this criterion for any $\rho$.

In the case of conservative estimates with high level $\alpha$, each term of~\cref{eq:VorobUncertainty} does not contribute equally to the expected distance in measure, as observed in~\cref{prop:boundT1error}. It is thus reasonable to consider the following criterion.

\begin{definition}[Type II criterion]
	Consider $Q_{n+q,\ConsLevel_{n}}$, the Vorob'ev quantile from $n+q$ evaluations with $\ConsLevel_{n}$, the conservative level obtained with $n$ evaluations. The
 \emph{Type II criterion} is defined as
\begin{align}
	\label{eq:T2criterion}
	J^{\textsc{t2}}_n(\newEval[q];\ConsLevel_{n}) &= \E_{n,\newEval[q]} \left[ H_{n+q}^{\textsc{t2}}(\ConsLevel_{n}) \right] \\ \nonumber
	&= \E_{n,\newEval[q]} \left[ G^{(2)}_n(Q_{n+q,\ConsLevel_{n}}) \right], \qquad \text{for }  \newEval[q] \in \inSpace^q.
\end{align}
\end{definition}

%
\Cref{prop:typeIIcrit},~\cref{sec:SeqProofs}, provides a fast-to-evaluate formula for~\eqref{eq:T2criterion}.

\begin{figure}
	\begin{minipage}{0.48\textwidth}
		\centering
		\includegraphics[width=\linewidth]{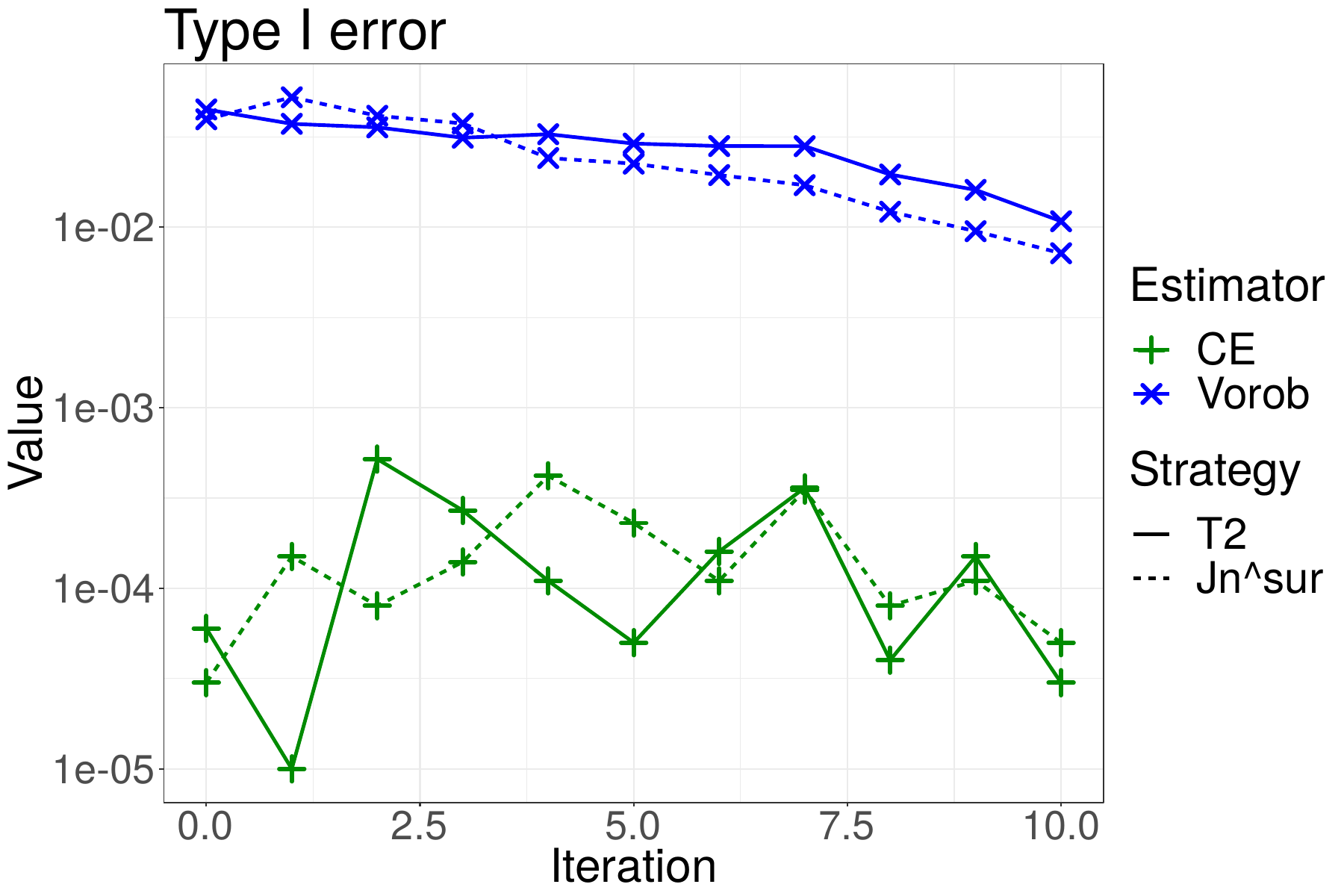}
		\subcaption{Empirical expected type I error computed on $100$ posterior realizations.}
		\label{fig:1dExSURtype1}
	\end{minipage}\hfill\hspace{0.05cm}
	\begin{minipage}{0.48\textwidth}
		\centering
		\includegraphics[width=\linewidth]{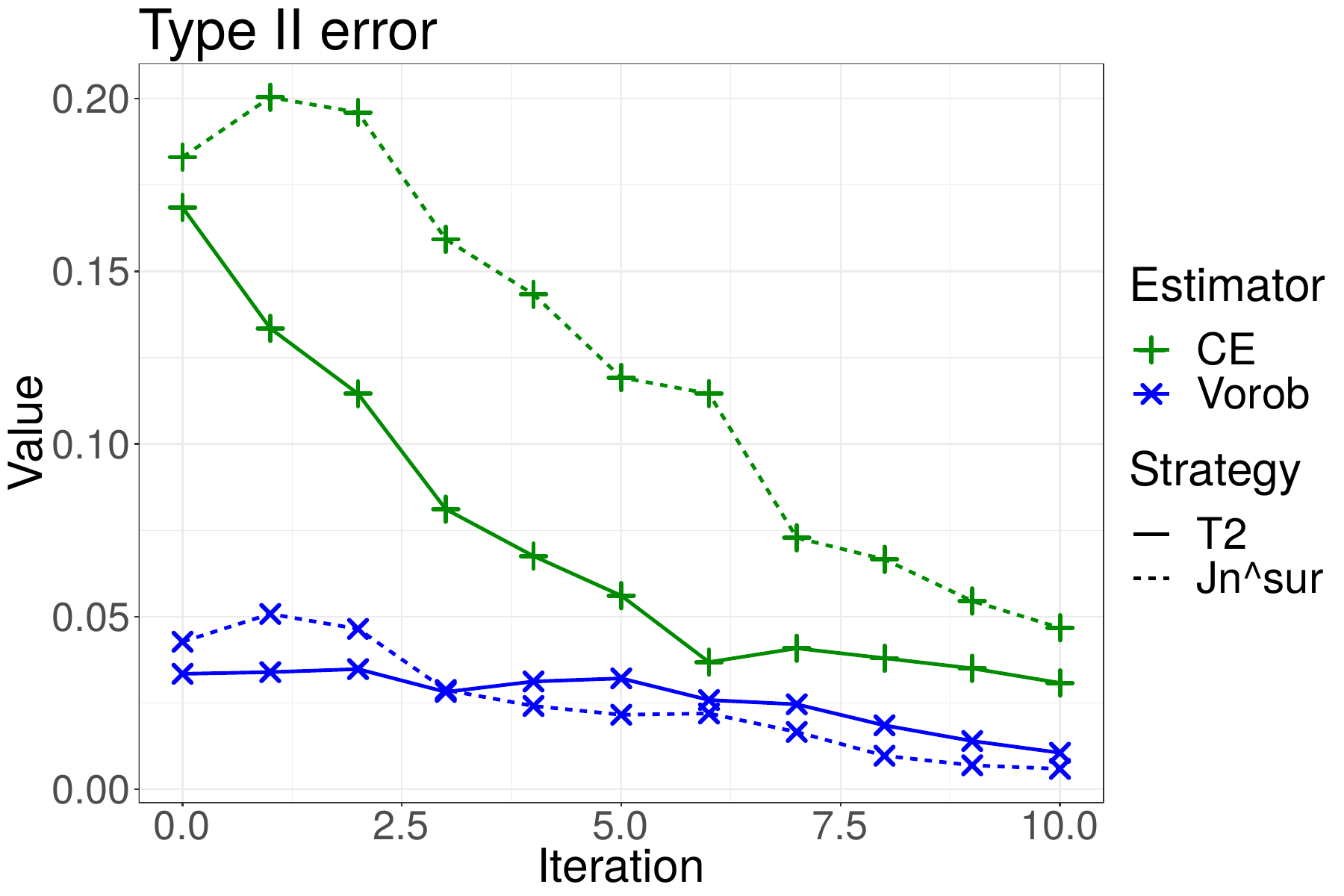}
		\subcaption{Empirical expected type II error computed on $100$ posterior realizations.}
		\label{fig:1dExSURtype2}	
	\end{minipage}
	\caption{Comparison of $J_n^{SUR}$ \citep{Bect.etal2012} and $J_n^{\textsc{t2}}$ on the example in~\cref{fig:1dExample}.}
	\label{fig:1dExSURcomparison}
\end{figure} 

The criteria $J_n, J_n^{\textsc{t2}}$ are implemented in this work with a plug-in approach for covariance hyper-parameters, i.e.\ at each step the hyper-parameters $\theta \in \Theta$ are estimated with maximum likelihood. A fully Bayesian approach would lead to a higher degree of conservativeness for the final estimate, as the hyper-parameter uncertainty would be accounted for. The formulae introduced in \cref{prop:VorobCrit,prop:typeIIcrit} could be adapted to a fully Bayesian approach, however 
their evaluation requires advanced Monte Carlo techniques \citep{Stroh2018} and it will be a future topic of research.

\Cref{fig:1dExSURcomparison} shows a comparison of estimated type I and type II errors obtained with strategy $J_n^{SUR}$ \citep[equation~(23)]{Bect.etal2012} and with strategy $J_n^{\textsc{t2}}$ in the experimental setting of~\cref{fig:1dExample}. Note how for the Vorob'ev expectation (blue $\times$ lines) the two strategies (dashed or solid lines) produce very similar results, however for conservative estimates (green $+$ lines), $J_n^{\textsc{t2}}$ (solid lines) reduces type II error faster than  $J_n^{SUR}$ (dashed lines).

\subsection{Implementation details}
\label{subsec:Implementation}


\Cref{prop:VorobCrit,prop:typeIIcrit}, in~\cref{sec:SeqProofs}, provide fast-to-evaluate expressions for the criteria, however their computation requires numerical approximations. See~\cref{sec:SeqProofs} for more details. New points are obtained by minimizing numerically the selected criterion, we use the genetic algorithm using derivatives of~\citet{rgenoud2011} to solve the optimization problem.

The strategies are implemented in the $\RprogLang$ programming language~\citep{Rcore} in the package \verb|KrigInv| \citep{Chevalier.etal2014a}. The function \verb|EGIparallel| in \verb|KrigInv| produces adaptive designs such as the one in~\cref{fig:1dExDoE10} by automatically optimizing the criterion~$J_n^{\textsc{t2}}$. \verb|KrigInv| interfaces with \verb|DiceKriging| \citep{Roustant.etal2012} for GP modeling, \verb|rgenoud| \citep{rgenoud2011} for the optimization routine and \verb|anMC| \citep{Azzimonti.etal2016b} for conservative estimates. 
See algorithm 1, in supplementary materials.



\section{Numerical benchmark for batch-sequential criteria}
\label{sec:TestCase}

\subsection{Noisy function evaluation scenarios}


\begin{table}
	\caption{\label{tab:ListGPnoise2}MC function evaluation scenarios, total cost $O(n_{MC} kq)$ fixed.}
	\centering
	\fbox{
		\begin{tabular}{ l  r  r  r  r  }
			$q$ & $\tau^2$ & $n_{MC}$ & $k$  & $n$ \\
			\hline
			$1$ & $6.25\cdot10^{-5}$ & $1.6\cdot10^4$ & $50$ & $50$ \\[2pt]
			$1$+$7$ & $5\cdot10^{-4}$ & $2\cdot10^3$ & $50$ & $400$ \\[2pt]
			$8$ & $5\cdot10^{-4}$ & $2\cdot10^3$ & $50$ & $400$ \\[2pt]
			$16$ & $10^{-3}$ & $10^3$ & $50$ & $800$ \\[2pt]
		\end{tabular}
	}
\end{table}

\

In this section we consider a synthetic numerical study that shows a practical use for batch-sequential criteria. The implementation of uncertainty quantification for expensive-to-evaluate computer experiments is often run on cloud computing platforms. When using such platforms, practitioners often have a fixed total computational budget which is determined, for example, by the money/time available to run experiments. Resources can be deployed in parallel by creating new computational nodes or sequentially by employing one node for longer time. Nodes are often virtual on such platforms, so there is no restriction on the number of parallel units available. 

Here we consider how to allocate resources in order to provide a conservative estimate and reduce the uncertainties for the set $\setOfInt$ in~\eqref{eq:setOfInt}. In our setting the evaluations of the function $f$ are approximated with Monte Carlo sampling, i.e.\ for $i=1, \ldots, n$, we obtain a value $z_i = \tfrac{1}{n_{MC}} \sum_{j=1}^{n_{MC}}(f(x_i) +\epsilon_{i,j})$, where $\epsilon_{i,j}$ are realizations of i.i.d. Gaussian random variables with zero mean and variance $\nu^2$. The number of Monte Carlo samples, $n_{MC}$, is fixed before the experiment starts and kept constant throughout. The observation model above can be written as $z_i = f(x_i) +\tau\epsilon_{i}$ with overall measurement noise $\tau^2 = \nu^2/n_{MC}$. The cost of one observation $z_i$ is proportional to $n_{MC}$ and for larger costs we achieve smaller variance $\tau^2$. Note that noise variance is assumed here homogeneous, i.e.\ $\nu^2$ does not depend on the location $x_i$. See~\cite{picheny2013quantile} for an example of online allocation applied to the problem of minimizing a noisy function with tunable precision. We consider an adaptive strategy with $k$ iterations of a batch sequential strategy that selects $q \geq 1$ new points for each iteration, i.e.\ $n=kq$ function evaluations overall. The total cost of the procedure is therefore $c_{tot} = O(k q n_{MC} ) = O(n n_{MC})$, where we assume that the costs of training the GP and of optimizing the criterion are negligible. Since $c_{tot}$ is fixed, then the choices of $n_{MC}$, $k$ and $q$ are linked. A larger $n_{MC}$ leads to more precise observations, however to a smaller overall number of evaluation $n$. 

We study four possible allocations of resources described in \cref{tab:ListGPnoise2} which range from a ``purely sequential'' strategy, where at each iteration all resources are used to reduce evaluation noise at a single input location, to a batch-sequential strategy where at each iteration q=16 different locations are explored with a high noise level. Note that the second strategy is a hybrid strategy where we add $1$ new evaluation with the criterion selected and $7$ others with a randomized LHS space-filling criterion. 

	\begin{table}
		\caption{\label{tab:ListStrategies}Strategies implemented in the test cases.}
		\centering
		\fbox{
			\begin{tabular}{ c  c  c }
				Strategy  & Criterion & Parameters \\
				\hline 
				Benchmark 1	& IMSE {\footnotesize(Integrated Mean Squared Error)} & \\[2pt]
				Benchmark 2	& tIMSE {\footnotesize(targeted IMSE)} & target=$t$ \\[2pt]
				A 			& $J_n(\cdot;\rho_n)$ & $\rho_n=0.5$ \\[2pt]
				B 			& $J_n(\cdot;\ConsLevel_n)$ & $\alpha=0.95$ \\[2pt]
				C			& $J^{\textsc{t2}}_n(\cdot;\ConsLevel_n)$ & $\alpha=0.95$ \\[2pt]
			\end{tabular}
		}
	\end{table}

We analyze the trade-off between batch size and noise level on a synthetic test case where we assume that the function $f$ is a realization of $(\xi_x)_{x \in \inSpace} \sim GP(\mean,\covkern)$ with constant mean function $\mean$ and Mat\'ern covariance kernel $\covkern$ with smoothness parameter $\nu=3/2$, variance $\sigma^2=1$ and lengthscales $\theta_i = 0.2$, $i=1,2$. The noise variance is described by the column $\tau^2$ in \cref{tab:ListGPnoise2} and depends on the specific scenario. The set to estimate is $\setOfInt = \{ x \in [0,1]^2 : f(x) \geq 1 \}$, an excursion above $t=1$. We use the volume on $[0,1]^2$ as the measure $\mu$. 
For each scenario we consider an initial DoE of size $n_{\text{init}} =3$ and a GP model with zero prior mean and Matern $3/2$ covariance kernel with hyperparameters $\sigma^2$ and $\theta_i$ known and set to the values specified above. We select the next function evaluation with the strategies listed in~\cref{tab:ListStrategies}, where we recall that the strategy IMSE chooses the next evaluation by minimizing the integrated mean squared error of the prediction, see, e.g.\ \citet{Sacks.etal1989} and $\tIMSE$ is the targeted IMSE strategy described in \citet{Picheny2010}. We run each strategy for the number of iteration specified in~\cref{tab:ListGPnoise2}. We consider $m_{\text{doe}} =10$ different initial DoE and, for each design, we replicate the procedure $10$ times with different values for $\xi_{\doe_{\text{init}}}$. 

	\begin{figure}
		\begin{minipage}{0.48\textwidth}
			\centering
			\includegraphics[width=1.05\linewidth]{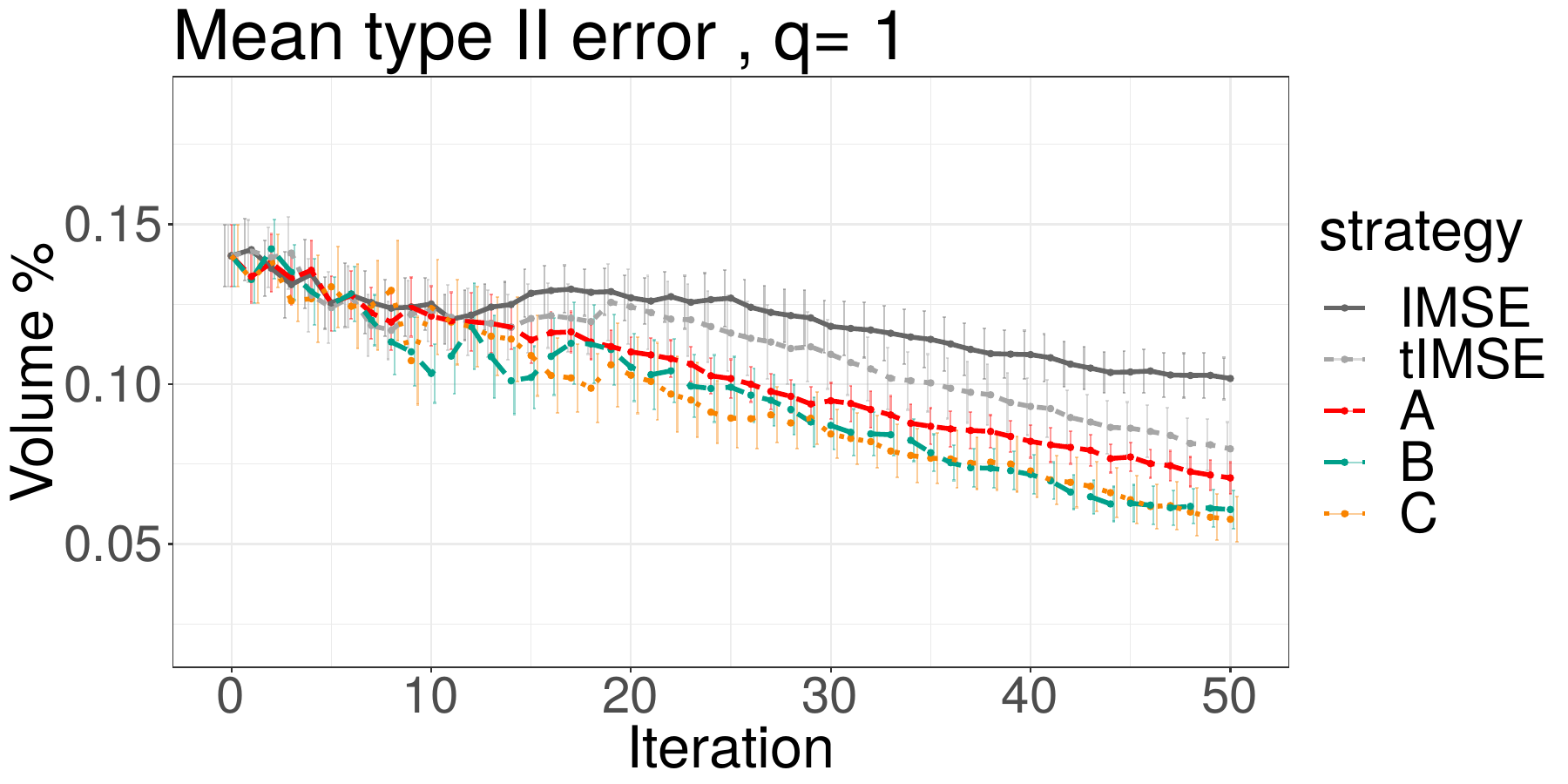}
			\subcaption{$q=1$.}
			\label{fig:gpNoiseT2q1}	
		\end{minipage} \hfill
		\begin{minipage}{0.48\textwidth}
			\centering
			\includegraphics[width=1.05\linewidth]{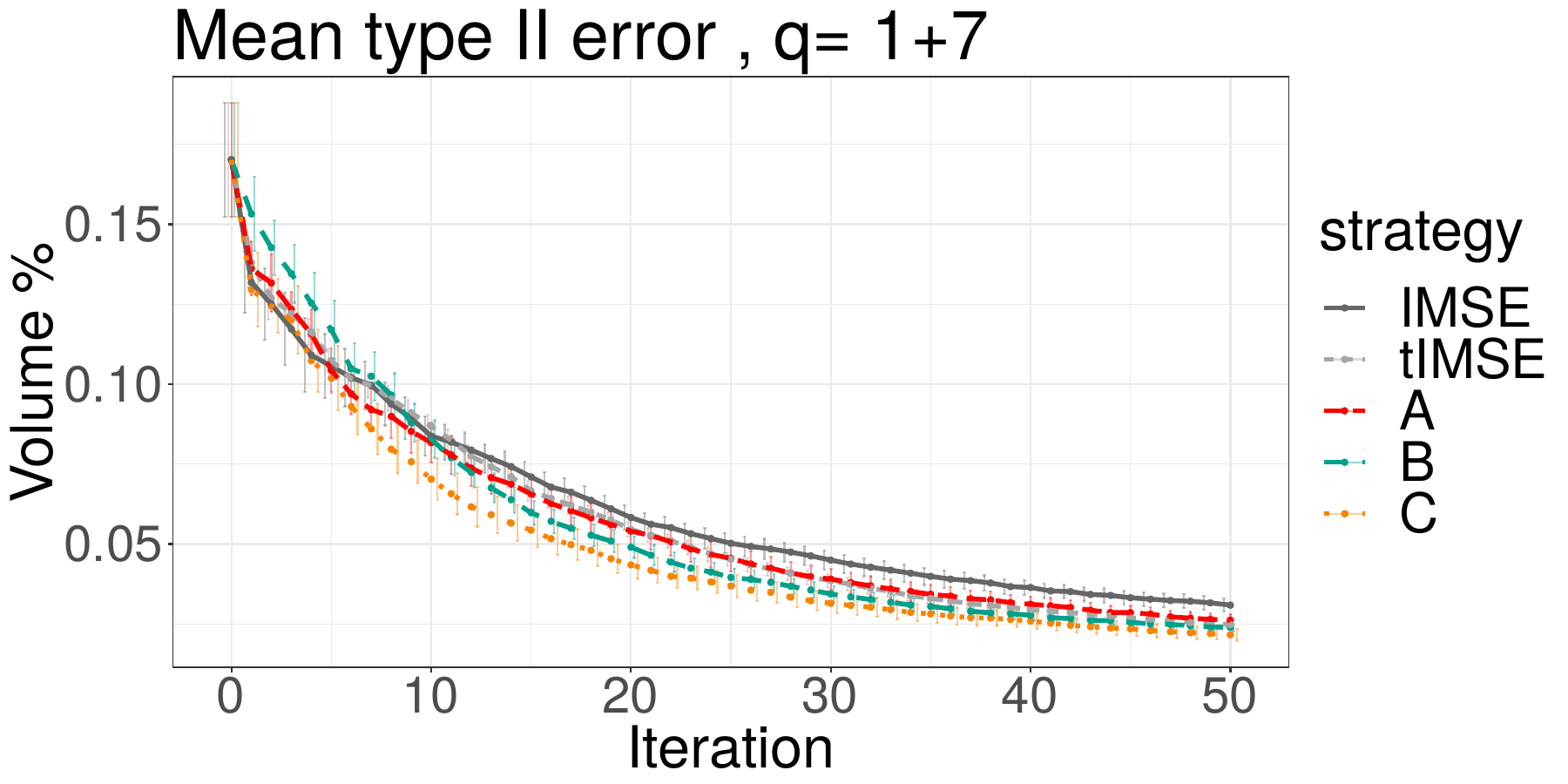}
			\subcaption{$q=1$+$7$. }
			\label{fig:gpNoiseT2q1+7}
		\end{minipage}\\
		\begin{minipage}{0.48\textwidth}
			\centering
			\includegraphics[width=1.05\linewidth]{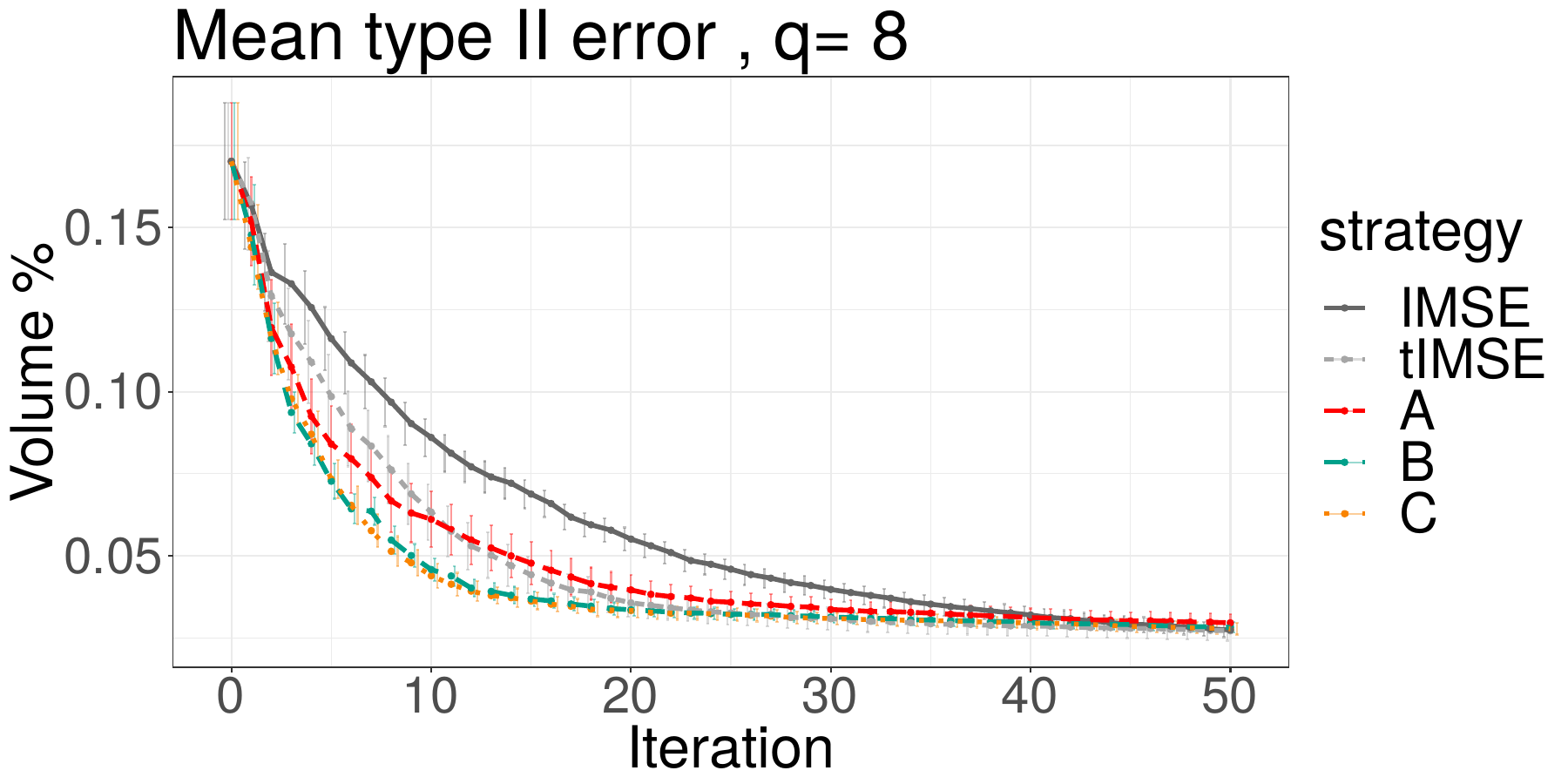}
			\subcaption{$q=8$. }
			\label{fig:gpNoiseT2q8}
		\end{minipage}\hfill
		\begin{minipage}{0.48\textwidth}
			\centering
			\includegraphics[width=1.05\linewidth]{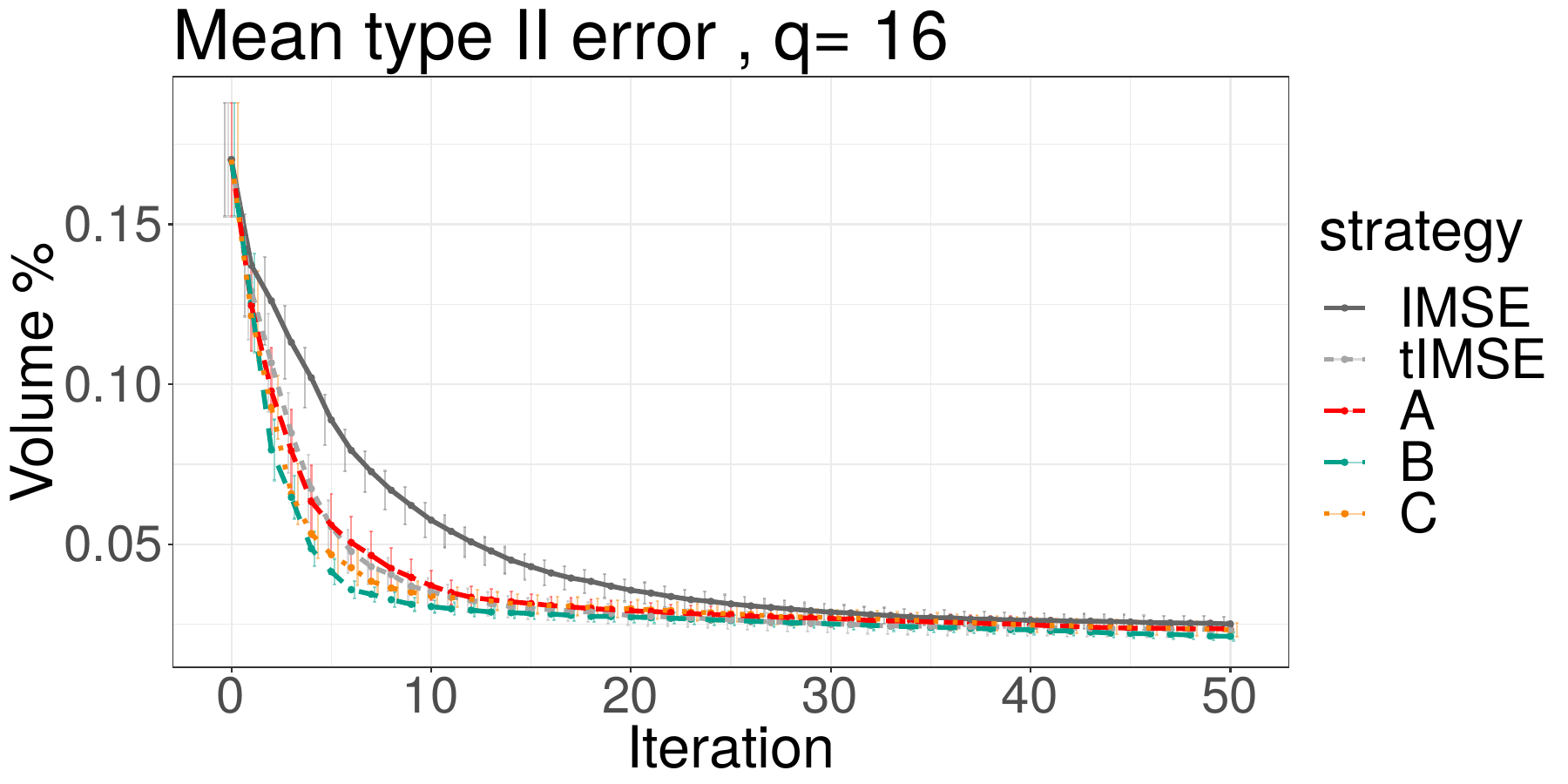}
			\subcaption{$q=16$. }
			\label{fig:gpNoiseT2q16}
		\end{minipage}
		\caption{Expected Type II error in different batch sequential scenarios.}
		\label{fig:gpNoiseT2}
	\end{figure}

	\Cref{fig:gpNoiseT2} shows a comparison of expected type II errors for the strategies listed in \cref{tab:ListStrategies} averaged over the $10$ replications and the $10$ initial DoEs. Scenario $q=1$ shows a faster decrease in type II error for strategies $B$ and $C$, however there are not enough function evaluations to reach convergence. For the other scenarios, the change in the median value of type II error is smaller than $1\%$ between successive iterations before the final iteration. Note that in \cref{fig:gpNoiseT2} (a), (c), (d) the differences between strategies are clear: strategies $C$ and $B$ are the fastest in reducing the error, followed by $A$ and $\tIMSE$; $\IMSE$ instead achieves the slowest error reduction. This reflects the fact that adaptive strategies tailored to the problem require fewer iterations to reduce type II error. In \cref{fig:gpNoiseT2} (b), the ranking between strategies is similar however the differences between strategies are less important. This is due to the effect of adding $7$ new input points with the same space-filling strategy independently of the criterion used to select the first point.

	\Cref{fig:gpNoiseT2} shows that, in this example, the parallel scenarios provide a much faster convergence for all strategies considered. 
	Another aspect to consider is wall\nobreakdash-clock time: under the assumption that Monte Carlo samples can be evaluated in parallel, then all scenarios require the same wall\nobreakdash-clock time. In some cases, however, the MC samples required for evaluating the function at one new input can be computed only on one computational node. Then the procedure with $q=1$ would be sequential so the wall\nobreakdash-clock time would be $O(n n_{MC})$, however, for $q>1$, each new input could be evaluated in parallel and the wall\nobreakdash-clock time would become $O(k n_{MC})$, where $k = \tfrac{n}{q}$. Therefore, in case of non-parallelizable MC computations, a greatly reduced wall-clock time would also be an additional benefit of batch sequential scenarios with $q>1$. 
	Note that here the noise is set relatively low in all scenarios. In high noise scenarios, where a strong trade-off between noise and number of evaluations is required, the situation is less clear. In supplementary material we present an example where the observations have higher noise variance. In such example, batch-sequential strategies still outperform sequential ones however the difference between scenarios is less pronounced. 
	
	
\subsection{Model-free comparison of strategies}

\begin{figure}
	\begin{minipage}{0.49\textwidth}
		\centering
		\includegraphics[width=1.05\linewidth]{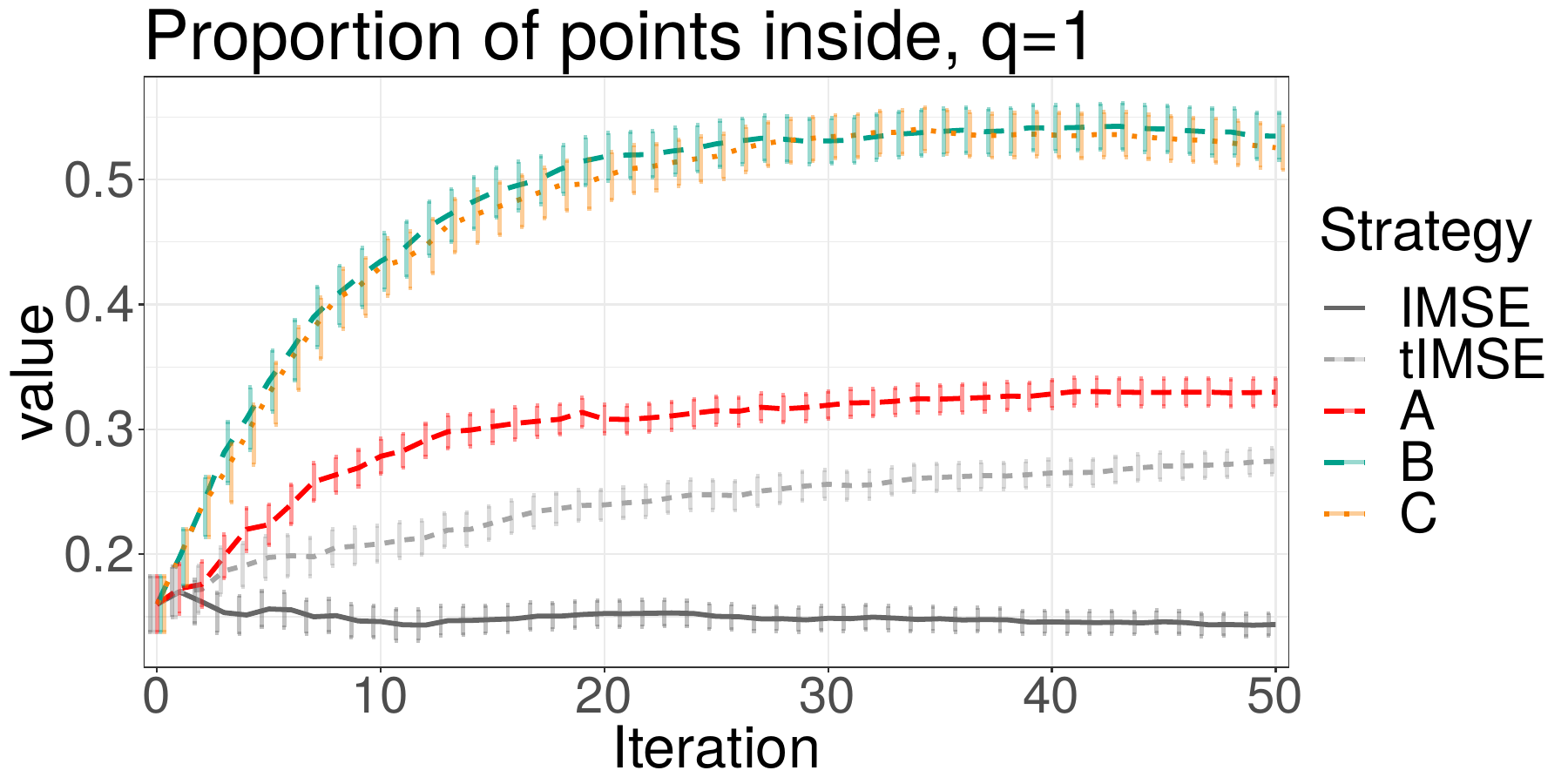}
		\subcaption{$q=1$.}
		\label{fig:pointsInq1}	
	\end{minipage} \hfill
	\begin{minipage}{0.48\textwidth}
		\centering
		\includegraphics[width=1.05\linewidth]{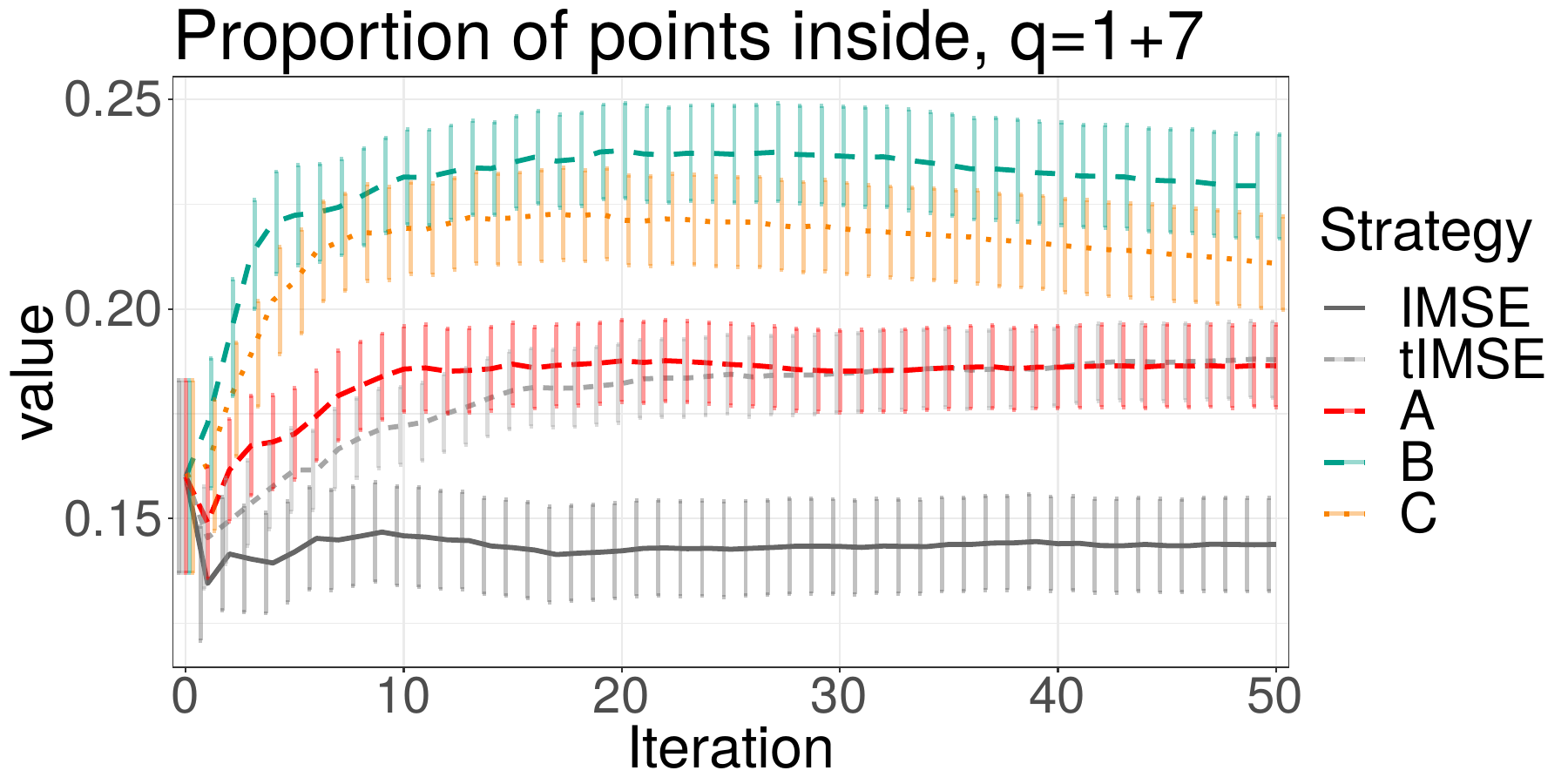}
		\subcaption{$q=1$+$7$.}
		\label{fig:pointsInq1+7}	
	\end{minipage} \\
	\begin{minipage}{0.48\textwidth}
		\centering
		\includegraphics[width=1.05\linewidth]{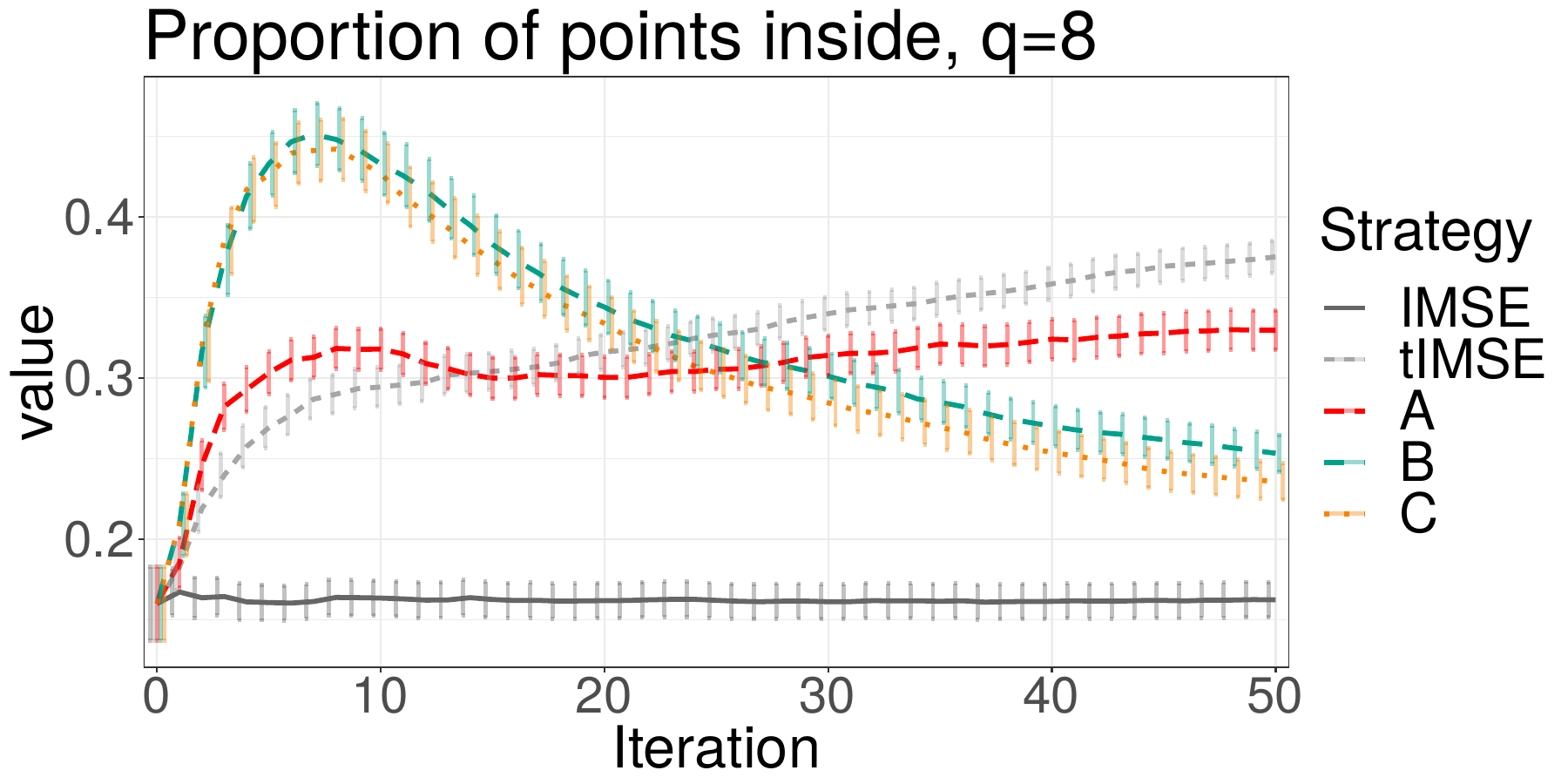}
		\subcaption{$q=8$.}
		\label{fig:pointsInq8}	
	\end{minipage} \hfill
	\begin{minipage}{0.48\textwidth}
		\centering
		\includegraphics[width=1.05\linewidth]{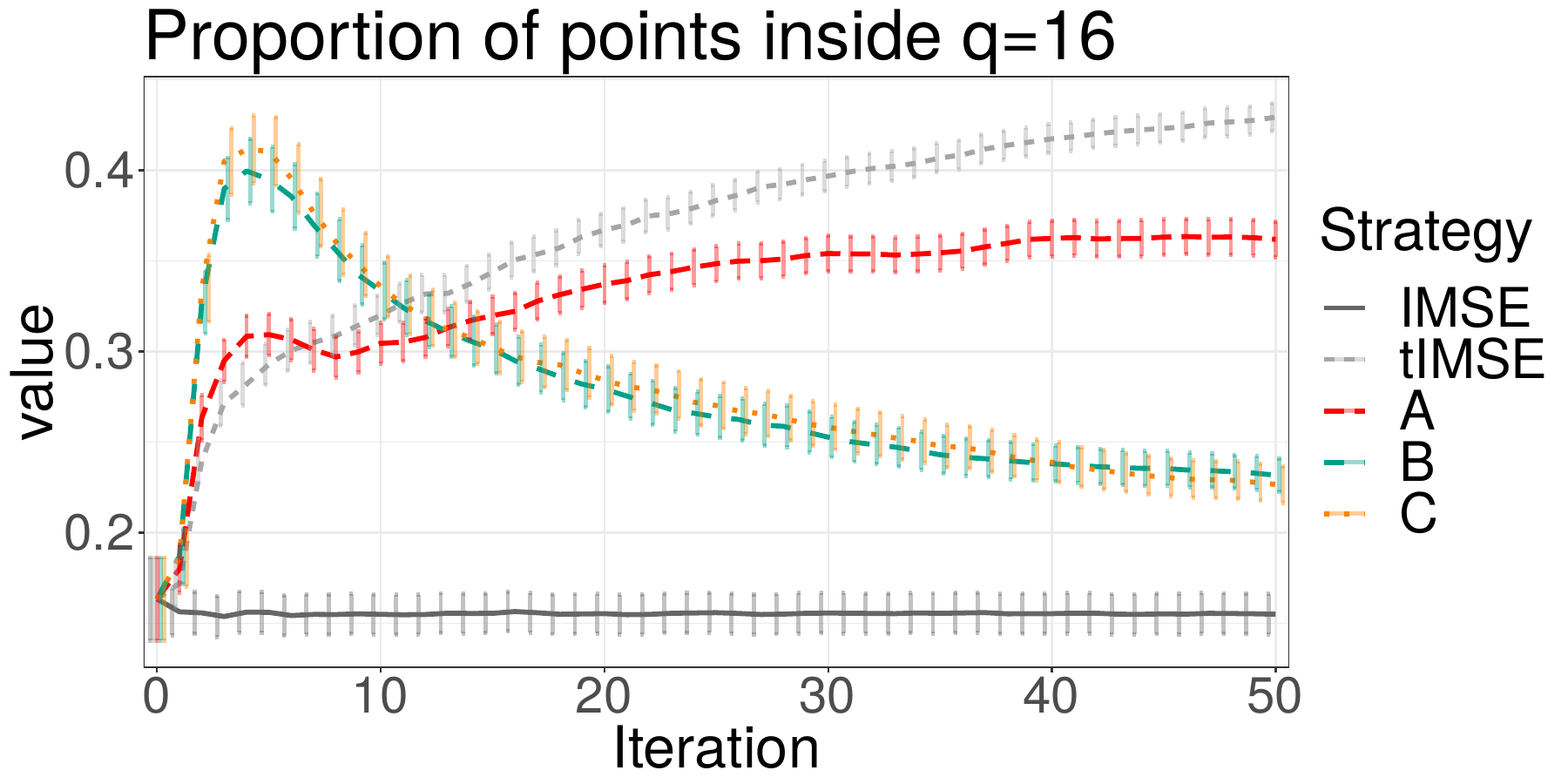}
		\subcaption{$q=16$.}
		\label{fig:pointsInq16}	
	\end{minipage}
	\caption{GP realizations. Average proportion of points inside the excursion region.}
	\label{fig:pointsInNoise}
\end{figure}

The metrics presented in the previous sections are based on the GP model. In this section we compare the strategies with a simpler metric independent from the underlying model.

We consider the number of evaluation points that are selected inside and outside the excursion set. At each iteration $i$, this quantity is computed as
$\frac{\# \{ j : z_j \geq t, \ j=1, \ldots, n_i \} }{n_i},$ 
where $n_i$ is the total number of points at iteration $i$ and $z_1, \ldots, z_{n_i}$ are the evaluations.  
\Cref{fig:pointsInNoise} shows the proportion of points inside the excursion set at each iteration for the three scenarios outlined in the previous section. Strategy $\IMSE$ is a space filling strategy therefore the proportion of points inside the excursion approximates the volume of excursion. In the scenario $q=1$, the strategies have not yet reached convergence, therefore both $B$ and $C$ tend to select more points inside the set than $A$ and $\tIMSE$ to consolidate the conservative estimation. The effect of the $7$ points chosen with a space-filling strategy in scenario $q=1+7$ is clear in \cref{fig:pointsInq1+7} where all strategies show proportions very close to $\IMSE$ which remain stable as the iteration number grows. On the other hand, \cref{fig:pointsInq8} and \cref{fig:pointsInq16} again show a fast increase in the proportion of points inside for strategies $B$ and $C$, however this is a transitory behavior and this proportion starts to decrease already after iterations $8$ and $10$ respectively. This highlights a tendency to explore the space by those strategies which was also verified visually by looking at the sequence of DoEs. Strategies $A$ and $\tIMSE$ instead tend to choose points around the boundary of the set therefore they initially choose fewer points inside the set and even after convergence they do not show an exploratory behavior.


\section{Reliability engineering test case}
\label{sec:IRSNtestCase}


In reliability engineering applications, the set $\setOfInt$ in~\cref{eq:setOfInt} often represents safe inputs for a system. In such settings, it is vital to avoid flagging unsafe regions as safe. 

\begin{figure}[t]
	\begin{minipage}{0.315\textwidth}
		\centering
		\includegraphics[width=1.15\linewidth]{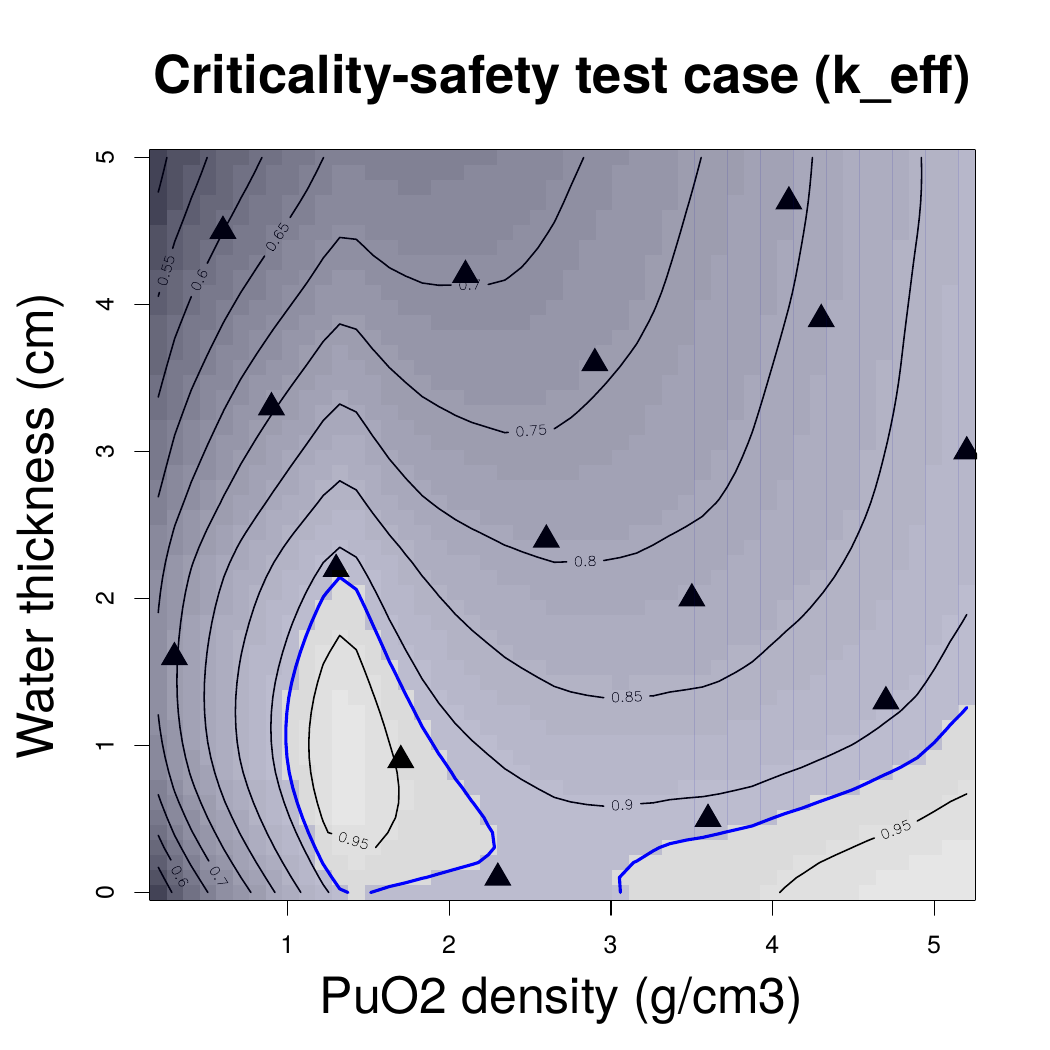}
		\subcaption{Function $\keff$, set of interest (shaded blue, $\volumeOf[\setOfIntKeff]$=$0.8816 \volumeOf[\inSpace]$) and initial DoE ($n=15$).}
		\label{fig:IRSNinitDoe}
	\end{minipage}\hfill\hspace{0.01\textwidth}
	\begin{minipage}{0.315\textwidth}
		\centering
		\includegraphics[width=1.15\linewidth]{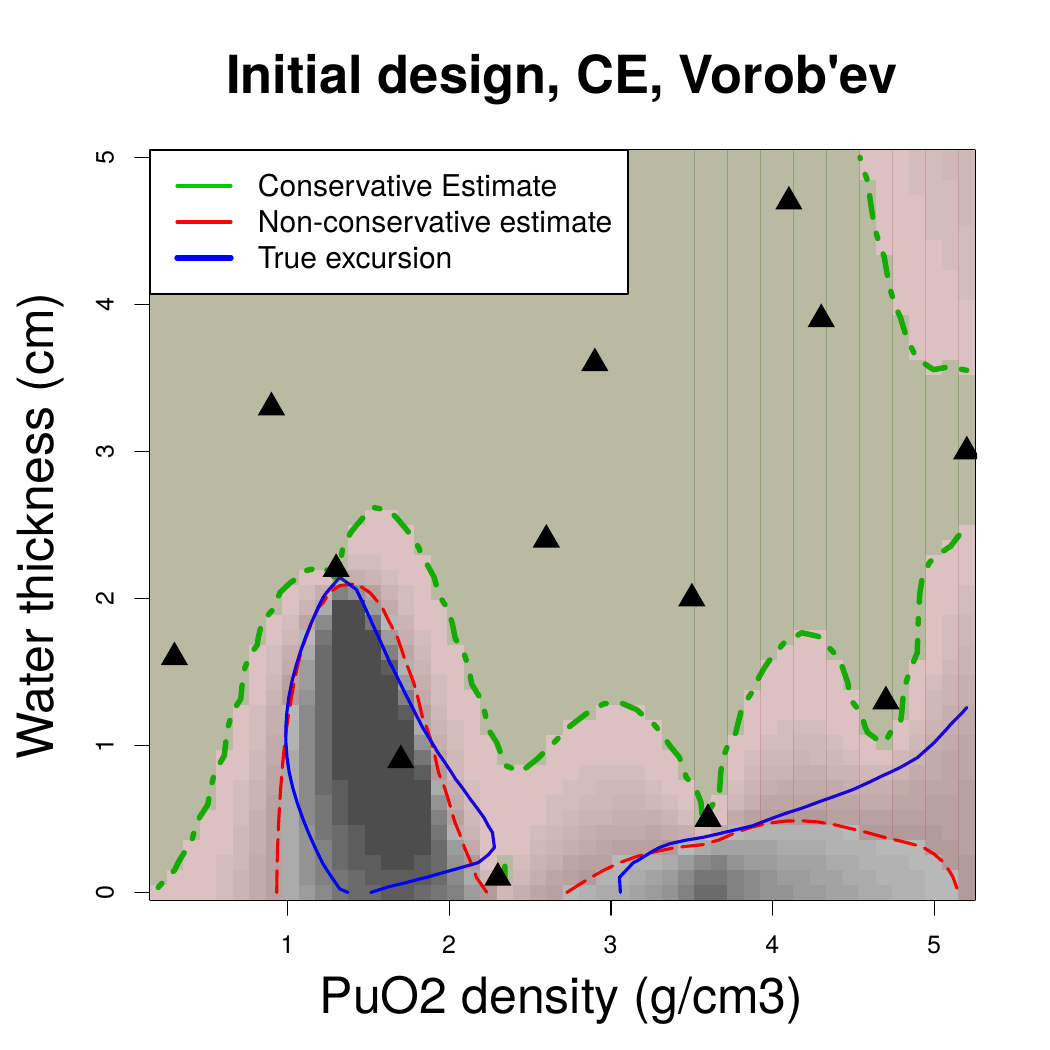}
		\subcaption{Conservative ($\alpha=0.95$, green) and non-conservative estimate (Vorob'ev expectation, red), initial DoE.}
		\label{fig:IRSNestComparisons}	
	\end{minipage} \hfill\hspace{0.01\textwidth}
	\begin{minipage}{0.315\textwidth}
		\centering
		\includegraphics[width=1.15\linewidth]{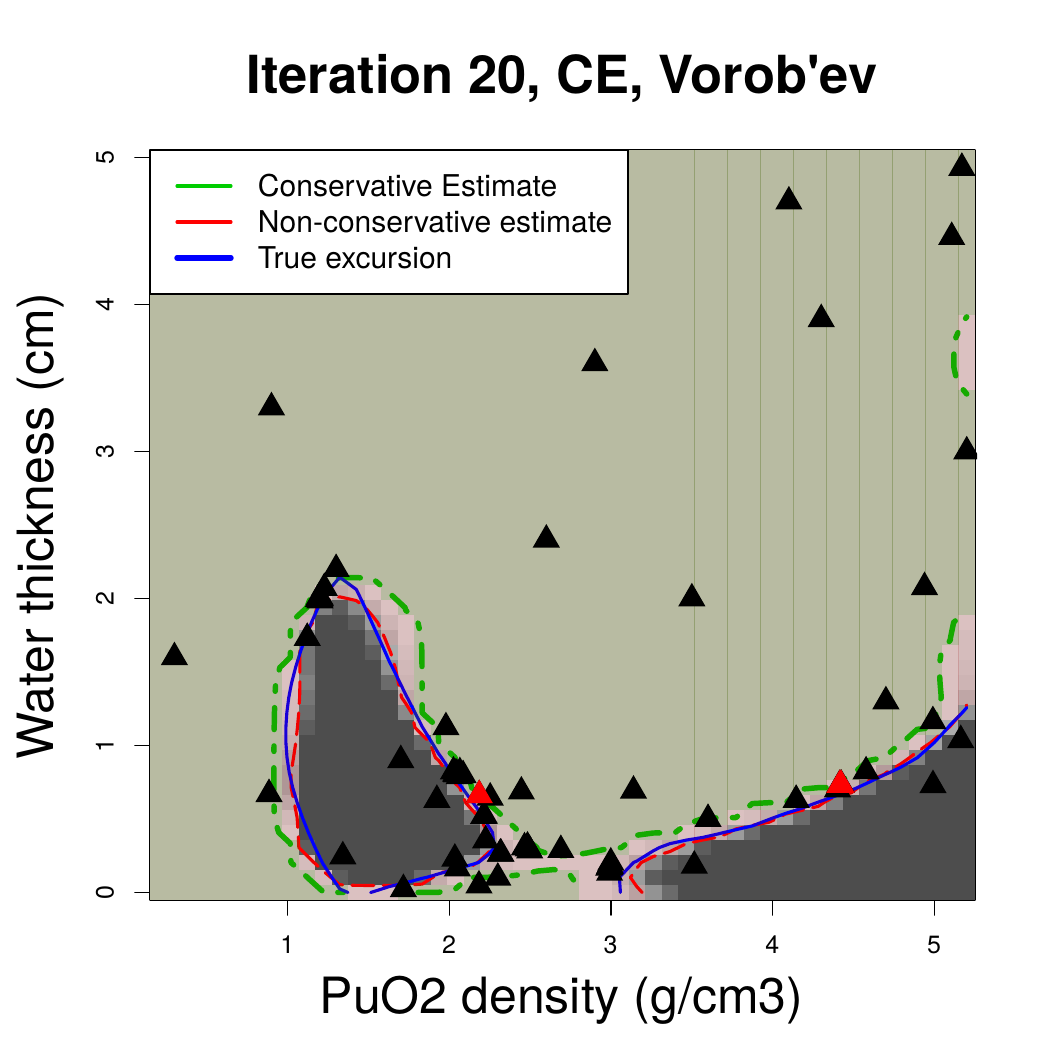}
		\subcaption{Conservative ($\alpha=0.95$, green) and non-conservative estimate (Vorob'ev expectation, red) after $75$ evaluations.}
		\label{fig:IRSNinitCE}	
	\end{minipage}
	\caption{Nuclear criticality safety test case. $\keff$ function (left), conservative and non-conservative estimates with $15$ (LHS design, middle) and $75$ (15+60 strategy $C$) evaluations.}
	\label{fig:IRSNinit}
\end{figure}

\Cref{fig:IRSNinit} shows an example of such reliability engineering applications: a test case from the French Institute for Radiological Protection and Nuclear Safety (IRSN). 
The problem concerns a nuclear storage facility and we are interested in estimating the set of parameters that lead to safe storage of the material. Since this is closely linked to the production of neutrons, the safety of a system is evaluated with the neutron multiplication factor produced by fissile materials, called $k$-effective or $\keff: \inSpace \rightarrow [0,+\infty)$. In our application $\inSpace=[0.2,5.2]\times[0,5]$ with the two parameters representing the fissile material density, $\PuOdens$, and the water thickness, $\WaterThick$. 
We are interested in the set of safe configurations
\begin{equation}
\setOfIntKeff = \{ (\PuOdens,\WaterThick) \in \inSpace : \keff(\PuOdens,\WaterThick)\leq 0.92 \},
\label{eq:setOfInt_IRSN}
\end{equation} 
where the threshold $t=0.92$ was chosen, for safety reasons, lower than the true critical case ($\keff=1.0$) where an uncontrolled chain reaction occurs. \Cref{fig:IRSNinitDoe} shows the set $\setOfIntKeff$ shaded in blue and the contour levels for the true function computed from evaluations over a $50\times50$ grid, used as ground truth.   The true data result from a MCMC simulation and have a heterogeneous noise variance. Here we consider the $\keff$ function in~\cref{fig:IRSNinit} obtained from $50\times 50$ evaluations of $\keff$ smoothed with a GP model that accounts for a prescribed value of noise variance provided by the simulator and considered as the true variance. 

We consider a GP model with covariance function from the Mat\'ern family and homogeneous noise variance estimated from the data. We choose the regularity parameter $\nu=5/2$ in order to represent the regularity of the underlying phenomenon. The initial DoE is a Latin hypercube sample design with $n_0=15$ function evaluations at the points plotted as triangles in~\cref{fig:IRSNinitDoe}. 
We consider the five strategies listed in~\cref{tab:ListStrategies} and we compare them on  $m_{\text{doe}}=10$ different initial DoEs of size $n_0=15$ obtained with the function \verb|optimumLHS| from the package \verb|lhs|~\citep{lhsRpackage} in $\RprogLang$.  

\Cref{fig:IRSNestComparisons} shows a conservative estimate at level $\alpha=0.95$ (shaded green) and a non conservative one (Vorob'ev expectation, shaded red) obtained from one of the $10$ DoEs, the true set $\setOfIntKeff$ is delimited in blue. 
\Cref{fig:IRSNinitCE} shows that, as more evaluations are available, conservative and non-conservative estimates both get closer to the true safe set. The estimates in this example are computed from $75$ function evaluations, where the last $60$ points were selected sequentially with strategy~$C$.

We now test how to adaptively reduce the uncertainty on the estimate with the strategies in~\cref{tab:ListStrategies}. We run $n=20$ iterations of each strategy and at each step we select a batch of $q=3$ new points where $\keff$ is evaluated. The covariance hyper-parameters are re-estimated at each iteration. 
The conservative estimates are computed with the Lebesgue measure $\measure$ on $\inSpace$. 

\begin{figure}
	\begin{minipage}{0.385\textwidth}
		\centering
		\includegraphics[width=1.1\linewidth]{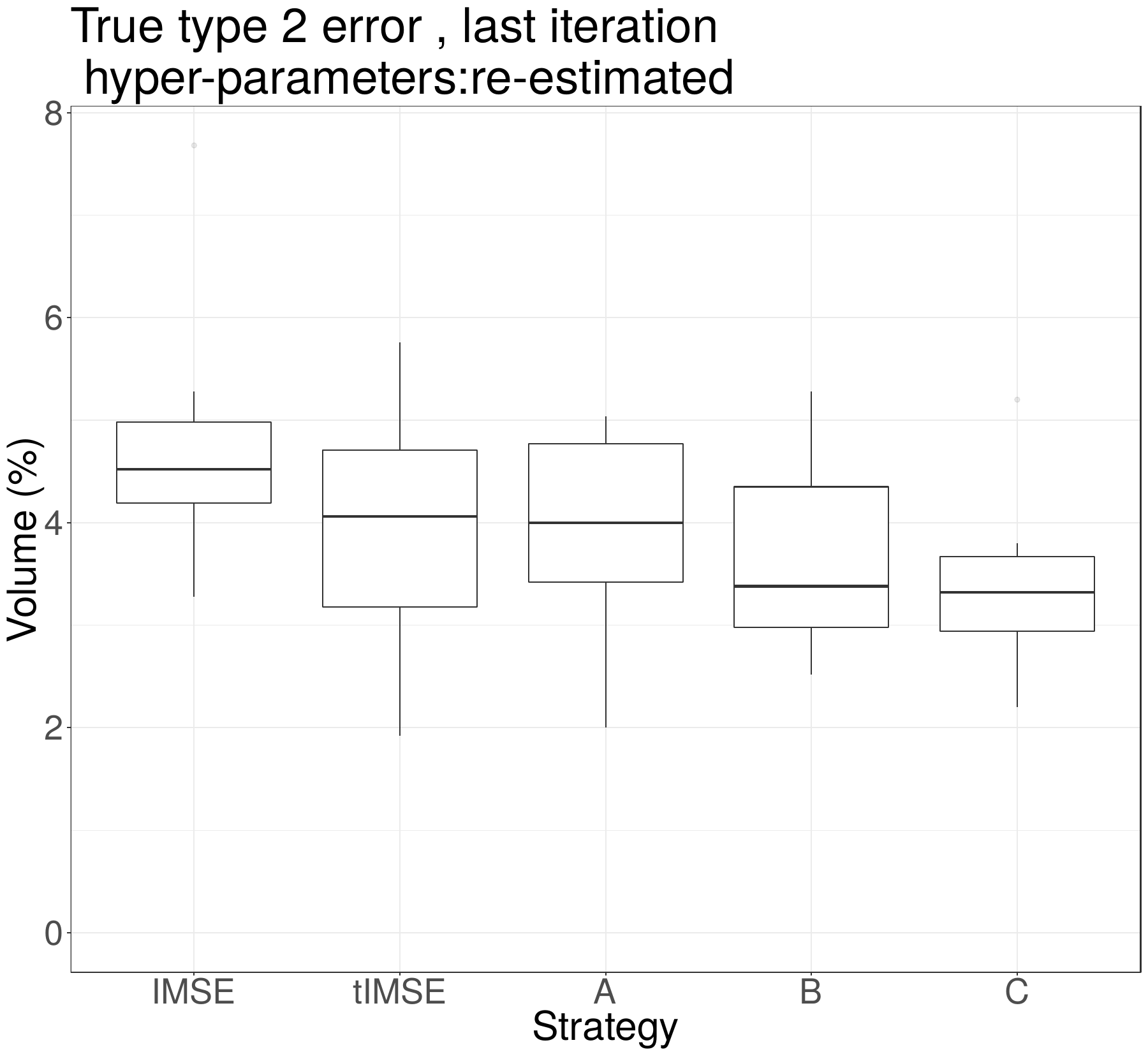}
		\subcaption{True type II error, last iteration.}
		\label{fig:IRSNrandDoEtypeII}	
	\end{minipage}\hspace{0.8cm}
	\begin{minipage}{0.585\textwidth}
		\centering
		\includegraphics[width=\linewidth]{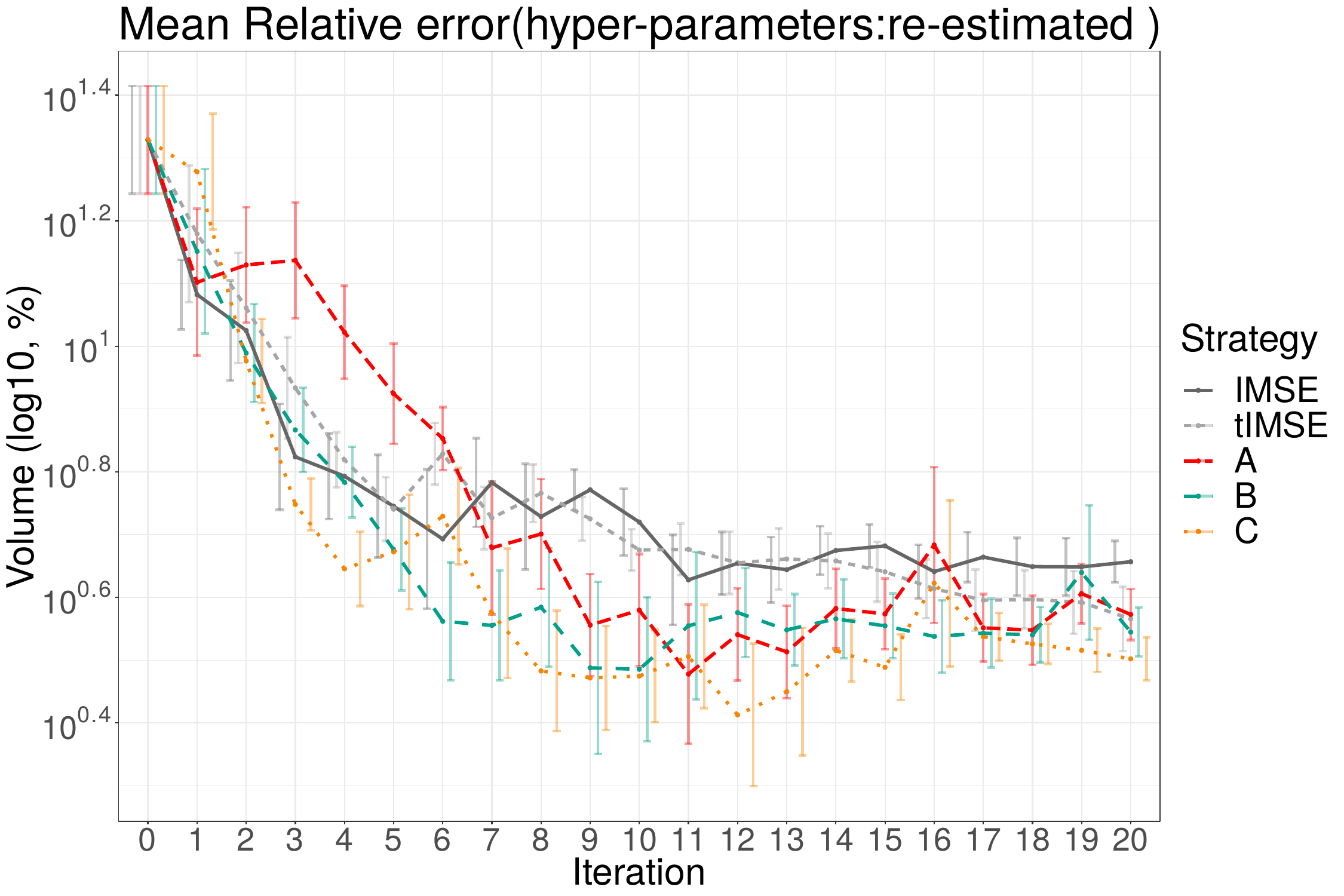}
		\subcaption{Relative volume error versus iteration number.} 
		\label{fig:IRSNvolErr}
	\end{minipage}
	
	\caption{Nuclear criticality safety test case, randomized initial DoEs.}
	\label{fig:IRSNrandDoE}
\end{figure}

\Cref{fig:IRSNrandDoEtypeII} shows the type II error (as percentage of the total measure of $\inSpace$) at the last iteration, i.e.\ after $75$ evaluations of the function, for each initial DoE and each strategy. Strategy $C$ achieves a median type II error $27\%$ lower than $\IMSE$. Strategy $B$ median type II error is $25\%$ lower than $\IMSE$ and strategy $A$'s $12\%$ lower than $\IMSE$. 


\Cref{fig:IRSNvolErr} shows the relative volume error as a function of the iteration number for strategies $\IMSE,\tIMSE,A,B,C$. The relative volume error is computed by comparing the conservative estimate with a ground truth for $\setOfInt$ obtained from evaluations of $\keff$ on a $50\times50$ grid. The volume of $\setOfInt$ computed with numerical integration from this grid of evaluations is $88.16\%$ of the total volume of the input space. All strategies show a strong decrease in relative volume error in the first $10$ iterations, i.e.\ until $30$ evaluations of $\keff$ are added, and strategies $B,C$ show the strongest decline in error in the first $5$ iterations. Overall, strategy $C$, the minimization of the expected type II error, seems to yield the best uncertainty reductions both in terms of relative volume error and type II error.

\section{Discussion}

	In this paper we introduced sequential uncertainty reduction strategies for conservative estimates. Such set estimates proved to be useful in a reliability engineering example, however they could be of interest in any situation where practitioners aim at controlling the overestimation of the set. 
	The estimator $\CE$, however, depends on the quality of the underlying GP model. Under the model, conservative estimates control, by definition, the false positive or type I error. If the GP model is not reliable then such estimates are not necessarily conservative. For a fixed model, increasing the level of confidence might mitigate this problem. We presented test cases with fixed $\alpha=0.95$, however testing different levels, e.g. $\alpha=0.99,0.995$, and comparing the results is a good practice. The computation of the estimator $\CE$ requires the approximation of the exceedance probability of a Gaussian process. This is currently achieved with a discrete approximation, however continuous approximations might be more effective.
	
	The sequential strategies proposed here provide a way to reduce the uncertainty on conservative estimates by adding new function evaluations. They were introduced with a homogeneous noise variance observation model, however as shown in~\cref{sec:SeqProofs}, the criteria implementations are available also in the heterogeneous noise variance case. Under such observation model, estimating the heteroskedastic noise variance structure can be challenging, see~\cite{Binois_etal2018} for more details. The numerical studies presented in the homogeneous and noise-free cases showed that adaptive strategies provide a better uncertainty reduction than generic strategies. In particular, strategy $C$, i.e.\ the criterion $J^{\textsc{t2}}_n(\cdot; \ConsLevel_n)$, is among the best criteria in terms of Type 2 uncertainty and relative volume error in all test cases. In this work we mainly focused on showing the differences between strategies with a-posteriori measures of uncertainty. Expected type I and II errors could also be used to provide stopping criteria for the sequential strategies. Further studies on those quantities could lead to a better understanding of their the limit behavior as $n$ increases.
	
	The strategies proposed in this work focus on reducing the uncertainty on conservative estimates. This objective does not necessarily lead to better overall models for the function or to good covariance hyper-parameters estimation. 
The sequential behavior of hyper-parameters maximum likelihood estimators under SUR strategies needs to be studied in more details and, in supplementary material, we report a small preliminary study on this aspect. On the other hand, a fully Bayesian approach, accounting for hyper-parameter uncertainty, could be used to strengthen the procedure's overall conservativeness.


\bigskip
\begin{center}
	{\large\bf SUPPLEMENTARY MATERIAL}
\end{center}

\begin{description}
	
	\item[Supplementary Materials:] additional test case, more details on the numerical benchmarks and theoretical complements to~\cref{subsec:UQconsEst} and \cref{subsec:CEvorobevQuant}. (pdf)
	
	\item[On-line code:] a git repository that allows (partial) reproducibility for the experiments in~\cref{sec:TestCase,sec:IRSNtestCase} and supplementary material sections~4 and~5 is available at \href{https://bitbucket.org/darioaz/supplemental_adoece}{supplemental\_adoece}. (git repository)
	
\end{description}

\bibliographystyle{apalike}
\bibliography{biblio}

\appendix

\section{Fast-to-evaluate formulae for sequential strategies}
\label{sec:SeqProofs}
In this section we prove two propositions that allow for the computation of the criteria in \cref{eq:VorobCriterion,eq:T2criterion}. We consider a more generic observation model than equation~\eqref{eq:obsModel}, where the noise variance $\tau$ is heterogeneous, i.e.\ $\tau(\cdot)$ is a function of $x\in \inSpace$.

First we extend the result in \citet{Chevalier2013} to any level $\rho_n$ which is a function of past $n$ observations.

\begin{proposition}[Criterion $J_n$]
	Consider $\setOfInt = \{x\in \inSpace : f(x) \in T \}$ with $T=[t, +\infty)$, where $t\in \R$ is a fixed threshold, then the criterion $J_n$ defined by~\cref{eq:VorobCriterion} can be expanded as
	\begin{align}
	\nonumber
	J_n(\mathbf{x}^{(q)};\rho_{n}) &= \E_{n,\newEval[q]} \left[\measure\left(\Gamma \Delta Q_{n+q,\rho_{n}} \right) \right] \\ \nonumber
	&= \int_\inSpace \bigg( 2\Phi_2 \left( \begin{pmatrix}
	a_{n+q}(u) \\ \Phi^{-1}(\rho_n) - a_{n+q}(u)
	\end{pmatrix}; \begin{pmatrix}
	1+\gamma_{n+q}(u) & -\gamma_{n+q}(u) \\ -\gamma_{n+q}(u) & \gamma_{n+q}(u)
	\end{pmatrix} \right) \\
	&-p_n(u) +\Phi\left( \frac{a_{n+q}(u) - \Phi^{-1}(\rho_n)}{\sqrt{\gamma_{n+q}(u)}} \right) \bigg)d\measureOf[u],
	\label{eq:critJV}
	\end{align}
	where 
	\begin{align}
	&a_{n+q}(u) = \frac{\mean_n(u) -t}{\sd_{n+q}(u)}, \ &\mathbf{b}_{n+q}(u) &= \frac{K^{-1}_q\covkern_n(\newEval[q],u)}{\sd_{n+q}(u)}, 
	\label{eq:abDef} \\ \nonumber
	&\gamma_{n+q}(u) = \mathbf{b}_{n+q}^T(u)K_q\mathbf{b}_{n+q}(u) \  &p_{n}(u)&= \Phi\left(\frac{\mean_n(u)-t}{\sd_n(u)}\right),  
	\ u \in \inSpace,
	\end{align}    
	with $\covkern_n(\newEval[q],u) = (\covkern_n(x_{n+1},u), \dots, \covkern_n(x_{n+q},u))^T$, $K_q = \covkern_n(\newEval[q],\newEval[q]) + \operatorname{diag}(\tau^2(\newEval[q]))$ is assumed invertible, $\covkern_n(\newEval[q],\newEval[q]) = [\covkern_n(x_{n+i},x_{n+j})]_{i,j=1,\dots,q}$, $\Phi_2(\cdot;\Sigma)$ is the cumulative distribution of the bivariate centered Normal with covariance matrix $\Sigma$ and $\Phi$ is the standard Normal cumulative distribution. 
	\label{prop:VorobCrit}	
\end{proposition}
\begin{proof}
	Recall that 
	\begin{equation}
	\E_{n,\newEval[q]}\left[\measureOf[\Gamma \Delta Q_{n+q,\rho_n}] \right] = \E_{n,\newEval[q]}[\underbrace{\measureOf[Q_{n+q,\rho_n} \setminus \Gamma] }_{= G^{(1)}_{n+q}(\rho_n) ]}] + \E_{n,\newEval[q]}[\underbrace{\measureOf[\Gamma \setminus Q_{n+q,\rho_n}]}_{= G^{(2)}_{n+q}(\rho_n) ]}] .
	\label{eq:Jnproof0}
	\end{equation}
	From the definitions of $G^{(1)}_{n+q},G^{(2)}_{n+q}$ and the law of total expectation we have
	\begin{align} \label{eq:Jnproof1}
	\E_{n,\newEval[q]}\left[G^{(2)}_{n+q}(\rho_n) \right] &= \int_{\inSpace}\E_n[ p_{n+q}(u)\mathds{1}_{\{p_{n+q}(u) < \rho_{n}\}} ] d\measureOf[u] \\ \label{eq:Jnproof2}
	\E_{n,\newEval[q]}\left[G^{(1)}_{n+q}(\rho_n) \right] &= \int_{\inSpace}\E_n[ \mathds{1}_{\{p_{n+q}(u) \geq \rho_{n}\}}(1-p_{n+q}(u)) ] d\measureOf[u] \\ \nonumber
	&= \int_{\inSpace}\left(\E_n[ \mathds{1}_{\{p_{n+q}(u) \geq \rho_{n}\}}] -\E_n[ \mathds{1}_{\{p_{n+q}(u) \geq \rho_{n}\}}p_{n+q}(u)) ] \right)d\measureOf[u] \\ \nonumber
	&= \int_{\inSpace}\left(\E_n[ \mathds{1}_{\{p_{n+q}(u) \geq \rho_{n}\}}] - p_n(u) \right)d\measureOf[u] + \E_{n,\newEval[q]}\left[G^{(2)}_{n+q}(\rho_n) \right]
	\end{align}
	Notice that, for each $x \in \inSpace $, the coverage function~$p_{n+q,\newEval[q]}$ can be written as
	\begin{equation}
	p_{n+q, \newEval[q]}(x) = \Phi \left( a_{n+q}(x) + \mathbf{b}_{n+q}^T Y_q \right),
	\label{eq:pnq}
	\end{equation}
	where $a_{n+q}, \mathbf{b}_{n+q}$ are defined in~\cref{eq:abDef} and $Y_q \sim N_q(0,K_q)$ is a $q$-dimensional normal random vector. The first part of~\cref{eq:Jnproof1} is 
	\begin{align}
	\nonumber
	\E_n[ \mathds{1}_{p_{n+q}(u) \geq \rho_{n}}] &= P_n(p_{n+q}(u) \geq \rho_n) = P_n(\mathbf{b}_{n+q}^T(u)Y_q \geq \Phi^{-1}(\rho_n)-a_{n+q}(u)) \\ \label{eq:JnProofPhi}
	&= \Phi\left( \frac{a_{n+q}(u) - \Phi^{-1}(\rho_n)}{\sqrt{\mathbf{b}_{n+q}^T(u) K_q\mathbf{b}_{n+q}(u)}} \right) 
	\end{align}
	where the second equality follows from~\cref{eq:pnq} and the third from $Y_q\sim N(0,K_q)$. Moreover
	\begin{align}
	\nonumber
	\E_n[ \mathds{1}_{\{p_{n+q}(u) < \rho_{n}\}} p_{n+q}(u)] &= \int \Phi \left( a_{n+q}(u) + \mathbf{b}_{n+q}^T(u) y \right)\mathds{1}_{\{b_{n+q}^T(u)y < \Phi^{-1}(\rho_n)-a_{n+q}(u)\}} \Psi(y) \\ \nonumber
	&=\int P(N_1 \leq a_{n+q}(u) + \mathbf{b}_{n+q}^T(u) y)\mathds{1}_{\{b_{n+q}^T(u)y < \Phi^{-1}(\rho_n)-a_{n+q}(u)\}} \Psi(y) \\ \nonumber
	&=\E\left[ P(N_1 \leq a_{n+q}(u) + \mathbf{b}_{n+q}^T y, \ \mathbf{b}_{n+q}^T(u)y < \Phi^{-1}(\rho_n)-a_{n+q}(u) )  \right] \\ 	\label{eq:Jnproof3}
	&=\Phi_2 \left( \begin{pmatrix}
	a_{n+q}(u) \\ \Phi^{-1}(\rho_n) - a_{n+q}(u)
	\end{pmatrix}; \begin{pmatrix}
	1+\gamma_{n+q}(u) & -\gamma_{n+q}(u) \\ -\gamma_{n+q}(u) & \gamma_{n+q}(u)
	\end{pmatrix} \right).
	\end{align}
	where $\Psi$ is the p.d.f. of $Y_q$, $N_1 \sim N(0,1)$. 
	By~\cref{eq:JnProofPhi,eq:Jnproof0,eq:Jnproof1,eq:Jnproof2,eq:Jnproof3} we obtain~\cref{eq:critJV}.

\end{proof}

We provide below a formulation for the SUR criterion $J^{\textsc{t2}}_n$ in~\cref{eq:T2criterion} which is fast-to-evaluate and allows for faster optimization. 

\begin{proposition}[Type II criterion]
	In the case $\setOfInt = \{x\in \inSpace : f(x) \in T \}$ with $T=[t, +\infty)$, where $t\in \R$ is a fixed threshold, the criterion $J^{\textsc{t2}}_n(\cdot;\ConsLevel_{n})$ can be expanded as
	\begin{align}
	\label{eq:critJC}
	J^{\textsc{t2}}_n(\newEval[q];\ConsLevel_{n}) &= \E_{n, \newEval[q]} \left[ G^{(2)}_n(Q_{n+q,\ConsLevel_{n}}) \right] \\ \nonumber
	&= \int_\inSpace \Phi_2 \left( \begin{pmatrix}
	a_{n+q}(u) \\ \Phi^{-1}(\ConsLevel_n) - a_{n+q}(u)
	\end{pmatrix}; \begin{pmatrix}
	1+\gamma_{n+q}(u) & -\gamma_{n+q}(u) \\ -\gamma_{n+q}(u) & \gamma_{n+q}(u)
	\end{pmatrix} \right) d\measureOf[u].
	\end{align}
	\label{prop:typeIIcrit}
\end{proposition}
\begin{proof}
	The proof follows from~\cref{eq:Jnproof1,eq:Jnproof3}.
\end{proof}

The evaluation of $J_n$ and $J^{\textsc{t2}}_n$ require the computation of an integral over $\inSpace$ with respect to $\measure$. The integral can be computed with an importance sampling Monte Carlo method as in~\cite{Chevalier.etal2014a} or by fixing the integration points with space filling designs, such as a Sobol' sequence or uniform sampling. If the dimension of $\inSpace$ is high, the region of interest for sampling could become very small with respect to $\inSpace$ and this would make simple Monte Carlo or importance sampling methods very inefficient. We did not observe this behavior in our experiments, however, in such cases sequential Monte Carlo (SMC) methods could provide better results. See, e.g.,~\citet{Bect.etal2017} and references therein.
We exploit the kriging update formulas \citep{chevalier2014corrected,emery2009kriging} for faster updates of the posterior mean and covariance when new evaluations are added. 
%

	\section{Properties of conservative estimates}
	\label{sec:PreliminariesAddRes}
	
	\subsection{Conservative estimates and confidence regions}
	\label{subsec:confidenceRegs}
	Consider an excursion set $\Gamma = \{ x \in \inSpace : Z_x \geq t \}$, $t \in \R$ and recall that a conservative estimate $Q_{\rho^*}$ for $\Gamma$ is chosen as the Vorob'ev quantile with $\rho^* \in \arg\max_{\rho \in [0,1]} \{ \mu(Q_\rho) : P(Q_\rho \subset \Gamma) \geq \alpha \}$. Since $Q_\rho \subset \Gamma \Leftrightarrow  \Gamma^C \subset Q_\rho^C$ and $\mu(Q_\rho^C) = \mu(\inSpace)-\mu(Q_\rho)$, then we have that $\rho^*$ is also the minimizer of $\rho \rightarrow \mu(Q_\rho^C)$ under the constraint $P(\Gamma^C \subset Q_\rho^C) \geq \alpha$. We can look at $Q_{\rho*}^C$ as a confidence region for $\Gamma^C$, in the sense that it is the smallest set that contains $\Gamma^C$ with a given probability. 
	
	In a reliability framework, if $\Gamma$ is the set of safe configurations, then by selecting a conservative estimate for the safe set, we are actually selecting a confidence region for the dangerous configurations.

	\subsection{Conservative estimates with Vorob'ev quantiles}
	\label{subsec:CEvorobevQuant}
	
	The conservative estimate definition in~\cref{eq:genericConservative} requires a family $\mathfrak{C}$ in which to search for the optimal set $\CE_{\alpha,\nNot}$. In practice, it is convenient to choose a parametric family indexed by a real parameter. Here we choose $\mathfrak{C} = \{Q_\rho : \rho \in [0,1]  \}$, i.e.\ the Vorob'ev quantiles. This is a nested family indexed by $\rho \in [0,1]$ where $Q_0 = \inSpace \in \mathfrak{C}$ and, for each $\rho_1 > \rho_2$,
	\begin{equation}
	Q_{\rho_1} \subset Q_{\rho_2}, \qquad Q_{\rho_1}, Q_{\rho_2} \in \mathfrak{C}.
	\label{eq:nestedSets}
	\end{equation}
	
	We now detail how to compute $\CE_{\alpha,\nNot}$ based on $\mathfrak{C}$, for a fixed $\alpha \in [0,1]$ from $n$ observations. For each $\rho \in [0,1]$, we define the function~$\psi_\Gamma: [0,1] \rightarrow [0,1]$ that associates to each $\rho$ the probability $\psi_\Gamma(\rho) := P_\nNot(Q_{\rho} \subset \Gamma)$. The function~$\psi_\Gamma$ is non decreasing due to the nested property in~\cref{eq:nestedSets}.
	Moreover, $\measure(Q_{\rho_1}) \leq \measure(Q_{\rho_2})$ for $\rho_1 \geq \rho_2$. The computation of $\CE_{\alpha,\nNot}$ amounts to finding the smallest $\rho=\ConsLevel_\nNot$ such that $\psi_\Gamma(\ConsLevel_\nNot) \geq \alpha$, which is achievable, for example, with a simple dichotomic search. 
	The procedure above is valid for any nested family of sets indexed by a real parameter, however, the Vorob'ev quantiles, in addition, have the following property.
	
	
	\begin{proposition}
		\label{prop:QuantileProperty}
		Consider a measure $\measure$ such that $\measureOf[\inSpace]<\infty$ and an arbitrary $\rho \in [0,1]$. A Vorob'ev quantile $Q_\rho$ minimizes the expected distance in measure with $\Gamma$ among all measurable $M$ such that $\measure(M)=\measure(Q_\rho)$.
	\end{proposition}
	
	\Cref{prop:QuantileProperty} is an extension of Theorem~2.3,~\citet{Molchanov2005} to a generic Vorob'ev quantile. As a consequence, a conservative estimate $\CE_{\alpha,\nNot} = Q_{\nNot,\ConsLevel_\nNot}$ computed with Vorob'ev quantiles minimizes the expected measure of false negatives ($\Gamma \setminus Q_{\nNot,\ConsLevel_\nNot}$) for fixed probability of false positives ($Q_{\nNot,\ConsLevel_\nNot} \setminus \Gamma$). 
	%
	%
	In general, the Vorob'ev quantile chosen for $\CE_{\alpha,\nNot}$ with this procedure is not the set $S$ with the largest measure satisfying the property $P(S \subset \Gamma) \geq \alpha$. See supplementary material for a counterexample. 
	

	\subsection{Proofs}
	
	In the following, let us denote by $(\Omega,\mathcal{F},P)$ a probability space.
	
	\begin{proof}[Proof of~\cref{prop:QuantileProperty}]
		We want to show that the set $Q_\rho$ satisfies%
		\begin{equation}
		\E\left[ \measure(Q_\rho \Delta \Gamma) \right] \leq \E\left[ \measure(M \Delta \Gamma) \right],
		\label{eq:vorQuantProperty}
		\end{equation}
		for each measurable set $M$ such that $\measure(M) = \measure(Q_\rho)$. Let us consider a measurable set $M$ such that $\measureOf[M]=\measureOf[Q_\rho]$. For each $\omega \in \Omega$, we have 
		\begin{align*}
		\measure(M \Delta \Gamma(\omega)) -  \measure(Q_\rho \Delta \Gamma(\omega)) &= 2\bigg(\measure(\Gamma(\omega) \cap (Q_\rho \setminus M)) - \measure(\Gamma(\omega) \cap (M \setminus Q_\rho)) \bigg) \\
		&+\measure(Q_\rho^C) - \measure(M^C).
		\end{align*}
		By applying the expectation on both sides, since $\measure(Q_\rho^C) = \measure(M^C)$, we obtain
		\begin{align*}
		\E\left[ \measure(M \Delta \Gamma) -  \measure(Q_\rho \Delta \Gamma) \right] &= \E\left[2\bigg(\measure(\Gamma \cap (Q_\rho \setminus M)) - \measure(\Gamma \cap (M \setminus Q_\rho)) \bigg) \right] \\
		&= 2\int_{Q_\rho \setminus M}{p_\Gamma(u) \dmeasure[u]} - 2\int_{M \setminus Q_\rho}{p_\Gamma(u) \dmeasure[u]}, 
		\end{align*}
		where the second equality comes from the definition of $Q_\rho$. Moreover, since $p_\Gamma(x) \geq \rho$ for $x \in Q_\rho \setminus M$ and $p_\Gamma(x) \leq \rho$ for $x \in M \setminus Q_\rho$ we have
		\begin{align*}
		2\left[\int_{Q_\rho \setminus M}{p_\Gamma(u) \dmeasure[u]} - \int_{M \setminus Q_\rho}{p_\Gamma(u) \dmeasure[u]} \right] &\geq 2\rho[\measure(Q_\rho\setminus M) - \measure(M \setminus Q_\rho) ] \\
		&= 2\rho[\measure(Q_\rho) - \measure(M) ] =0,
		\end{align*}
		which shows that $Q_\rho$ verifies equation~\cref{eq:vorQuantProperty}.
	\end{proof}

	\begin{proof}[Proof of~\cref{prop:boundT1error}]
		Notice that for all $\omega \in \Omega$ such that $Q_{\nNot,\ConsLevel_\nNot} \subset \Gamma(\omega)$, we have $G^{(1)}_\nNot(\omega) = 0$. By applying the law of total expectation we obtain
		\begin{align*}
		\E_\nNot[ G^{(1)}_\nNot ] &= \E_\nNot[G^{(1)}_\nNot \mid Q_{\nNot,\ConsLevel_\nNot} \subset \Gamma]P(Q_{\nNot,\ConsLevel_\nNot} \subset \Gamma) \\ &+ \E_\nNot[ G^{(1)}_\nNot \mid Q_{\nNot,\ConsLevel_\nNot} \setminus \Gamma \neq \emptyset]( 1- P(Q_{\nNot,\ConsLevel_\nNot} \subset \Gamma) ) \\
		&\leq  0 + \E_\nNot[ G^{(1)}_\nNot \mid Q_{\nNot,\ConsLevel_\nNot} \setminus \Gamma \neq \emptyset]( 1- \alpha ) \leq \measureOf[Q_{\nNot,\ConsLevel_\nNot}](1-\alpha).
		\end{align*}
	\end{proof}

\end{document}